\documentclass{ar-1col}
\pdfoutput=1
\usepackage{rotating}%
\usepackage{aas_macros} 
\usepackage[numbers]{natbib}
\jname{Xxxx. Xxx. Xxx. Xxx.}
\jvol{00}
\jyear{2015}
\doi{10.1146/((please add article doi))}

\begin{document}

\markboth{Funk}{Space- and ground-based Gamma-ray Astrophysics}

\title{Space- and Ground-Based Gamma-Ray Astrophysics}

\author{Stefan Funk $^1$, $^2$%
\affil{$^1$Erlangen Centre for Astroparticle Physics,
  Friedrich-Alexander-Universit{\"a}t Erlangen-N{\"u}rnberg,
  Erwin-Rommel-Str. 1, D-91058 Erlangen, Germany} 
\affil{$^2$Kavli Institute for Particle Astrophysics and Cosmology,
  Department of Physics and SLAC National Accelerator Laboratory,
  Stanford University, Stanford, CA 94305, USA}
}

\begin{abstract}
  In recent years, observational $\gamma$-ray astronomy has seen a
  remarkable range of exciting new results in the high-energy and
  very-high energy regimes. Coupled with extensive theoretical and
  phenomenological studies of non-thermal processes in the Universe
  these observations have provided a deep insight into a number of
  fundamental problems of high energy astrophysics and astroparticle
  physics. Although the main motivations of $\gamma$-ray astronomy
  remain unchanged, recent observational results have contributed
  significantly towards our understanding of many related
  phenomena. This article aims to review the most important results in
  the young and rapidly developing field of $\gamma$-ray astrophysics.
\end{abstract}


\maketitle


\section{INTRODUCTION}
$\gamma$-ray astrophysics has seen a remarkable progress in recent
years. High-energy observations of the Universe now represent a firmly
established discipline in modern Astrophysics as they allow for
studies of a broad range of topics related to the non-thermal
Universe. The acceleration, propagation, and $\gamma$-ray radiation of
relativistic particles is by now detected on a vast range of
astronomical scales: from compact objects like neutron stars and solar
mass black holes to cosmological scales such as other Galaxies and
active galactic nuclei. At the same time, $\gamma$-ray astronomy can
be considered as a discipline in its own right with interesting links
to Cosmology and Particle Physics. $\Gamma$-ray photons travel in
straight lines and therefore allow for an accurate determination of
their origin. While $\gamma$ rays often denote photons above 100 keV,
(i.e.\ frequencies $\geq 10^{19}$ Hz), the focus of this review will
be on astrophysical gamma rays above $\sim 10$ MeV ($10^7$ eV), the
region where pair production into electron-positron pairs is the
dominant interaction mechanism for photons. The highest-energy photons
currently detected are in the range of 100~TeV ($10^{14}$
eV)~\citep{HESS:rxj1713p3}.  Consequently, this review will cover an
energy range of about 7 decades in energy. Recent results were
obtained with both space-based instruments such as the {\emph{Fermi}}
Large Area Telescope (LAT)~\citep{LATPaper} and the AGILE
satellite~\citep{AGILE} and with ground-based observatories such as
H.E.S.S.~\citep{HESS:phaseII}, MAGIC~\citep{MAGICII:status}, and
VERITAS~\citep{VERITAS:status}.  These observations have changed and
deepened our understanding of energetic processes in the Universe and
make this review timely.

Gamma rays are the highest-energy form of electromagnetic radiation.
As such, the origins of gamma-ray astronomy are tightly connected to
the scientific investigations into the origin of cosmic rays (CRs) --
the highest energy particles in the Universe. First estimates for
fluxes of cosmic gamma-ray sources (as a result of CR interactions)
were given in a seminal paper by Morrison~\citep{Morrison1958}. These
estimates turned out to be significantly overestimating the real
fluxes, but they spawned a series of rocket and balloon experiments
looking for gamma-ray emission from the sources of CRs. Space-based
missions were used in those early days because Earth's atmosphere
blocks gamma rays. Space-based observations in the early to mid-1960s
mainly through satellites of gamma rays in the energy region around
100 MeV led to a number of breakthroughs in the study of the
high-energy Universe. These came as first detection of gamma rays from
space by Explorer XI~\citep{KraushaarClark1962} (mostly produced in
interactions of CRs with the Earth atmosphere), with the first
detection of gamma-ray bursts using the Vela
satellites~\citep{VelaSatellites}, and finally with the first
detection of gamma rays from the interaction of CRs with interstellar
material of the Milky way by OSO-3~\citep{Kraushaar1972}. Subsequent
years and decades saw significant improvement in the technology used
in space-based detectors and orders of magnitude improvement were
reached by employing instruments such as SAS-2~\citep{Fichtel1975},
COS-B~\citep{Bigniami}, and EGRET~\citep{EGRET3:EG}.

During these times, also first ideas for observing gamma rays from the
ground were developed. The ideas were based on the fact that in the
interaction of gamma rays with the atmosphere the primary gamma ray
gets lost, but secondary products are created that can be detected on
the ground and that can be used to infer the properties of the
astrophysical gamma rays. Already in the early
1950s~\citep{GalbraithJelley} detected Cherenkov (optical) emission
from atmospheric showers. The difficulty in the early days of this
technique was how to suppress the much more numerous proton showers,
that also generate Cherenkov light.  Distinguishing gamma-ray induced
showers from proton-induced showers based on their Cherenkov light
turned out to be a significant challenge, although the basic idea to
make use of the difference in shower development in the atmosphere
(and thus in the image shape in the detector) was already suggested in
the early 1960s~\citep{JelleyPorter1963}. However, it took a heroic
effort mostly by groups in the US and in
France to develop the technique to the point where the
first astrophysical source (the Crab Nebula) could be detected at TeV
gamma-ray energies from the ground~\citep{Whipple:Crab}. The main
technical breakthrough was to use a pixelated camera to exploit the
difference in shape of the atmospheric shower between gamma rays and
protons and thereby suppress the background. The next big step forward
came in the use of stereoscopy, i.e. by connecting several imaging
Cherenkov telescopes to view the shower from different angles and
further suppress the background~\citep{HEGRA:acts}. By now all major
installations (H.E.S.S., MAGIC, VERITAS) employ a system of more than
one telescope with pixelated cameras that cover a field of view of
typically 5$^\circ$ diameter and $\sim 1000-2000$ pixels in each
camera~\citep{HintonHofmann}.

\begin{figure}[!htb]
\centering
\includegraphics[width=0.73\textwidth]{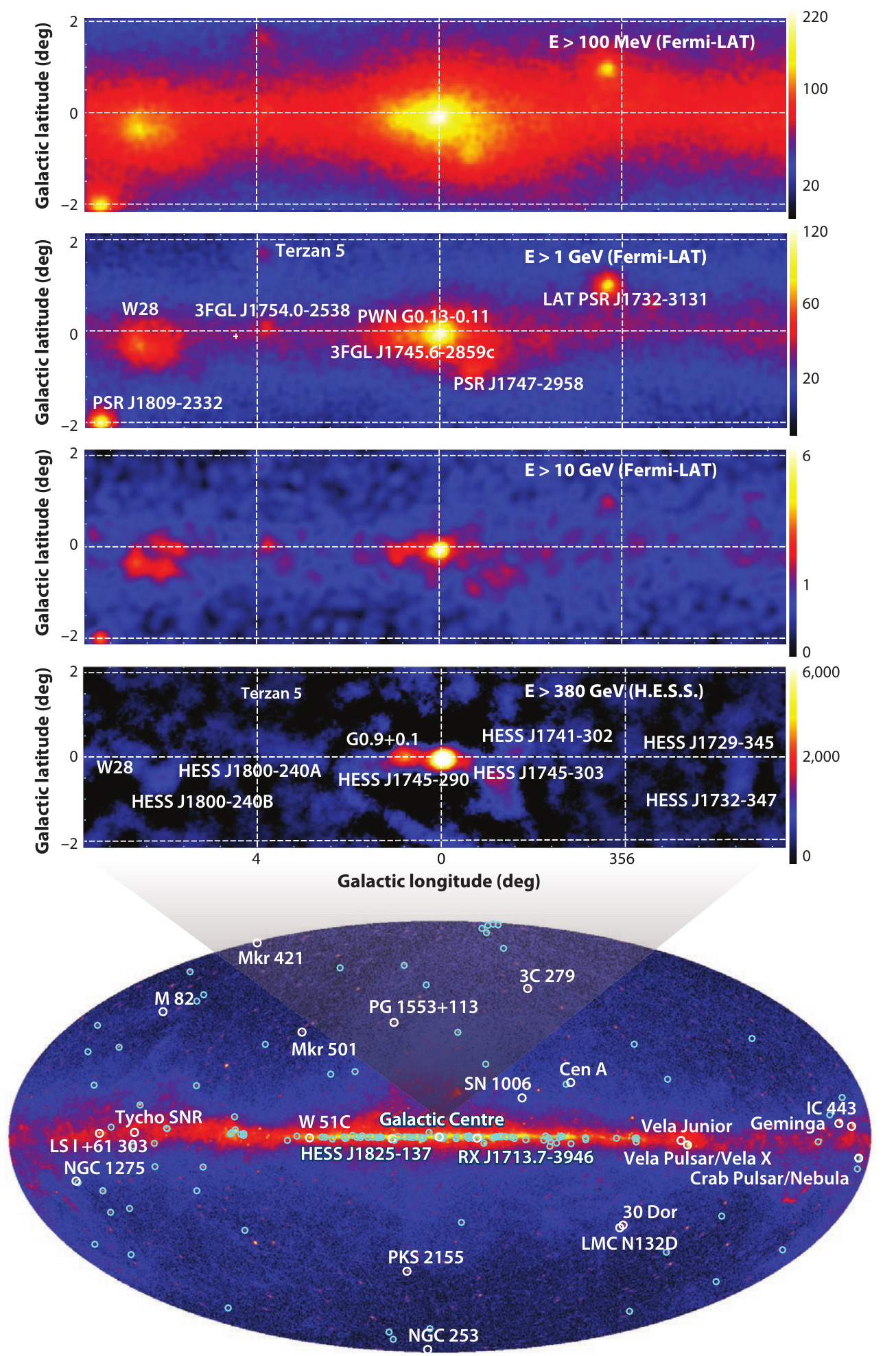}
\caption{(Bottom) The high-energy $\gamma$-ray sky. The colors
  correspond to intensity per pixel (i.e., counts divided by exposure
  and solid angle) for Pass 8 Source class (PSF 3 event class—the
  highest angular resolution event class), for $\gamma$-rays above 1
  GeV. The positions of discrete TeV $\gamma$-ray sources detected
  with ground-based instruments are marked by cyan circles or white
  circles taken from “TeVCat (http://tevcat.uchicago.edu)”. (Top)
  Zoomed-in view into the Galactic center region ($|l| < 8^{\circ}$,
  $|b| < 2^{\circ}$; corresponds to a size of $1,100 \times 280$ pc) for
  different minimum energies. Several prominent sources from the 3FGL
  catalog are marked in the second panel. Note that a significant
  number of sources from 3FGL are not marked. The fourth panel shows a
  H.E.S.S. excess map from the Galactic plane survey as published for
  2004 data (24) above 380 GeV. The TeV-detected sources from TeVCat
  are marked, although certain areas of this region received a
  significantly larger number of observations in subsequent years.}
\label{fig::FermiSky}
\end{figure}

Our current view on the gamma-ray Universe is summarised in Figure
~\ref{fig::FermiSky}, showing the gamma-ray sky above 1 GeV in
Galactic coordinates as observed by the {\emph{Fermi}}-LAT. As can be seen, the
sky is composed of a broad band of diffuse emission along the Galactic
plane, stemming from the interaction of the pool of Galactic CRs with
interstellar gas and dust and in addition a collection of
individual sources. The {\emph{Fermi-LAT}} has provided significant
new insights into the high-energy gamma-ray sky. Important discoveries
include the observation of the pion-decay cutoff in the spectra of
supernova remnants (SNRs), and multi-GeV components of gamma-ray
bursts (GRBs), {\it etc} \citep[see e.g. a review article
by][]{ReviewDermer}. The LAT also significantly enhanced our knowledge
on the diffuse gamma-ray backgrounds. In particular, {\it Fermi}-LAT
has extended the range of observations of the diffuse emission of the
Galactic Disk and the isotropic (extragalactic) gamma-ray background
to several hundreds of GeV. Since the energy spectrum of the Galactic
diffuse emission is much softer (meaning that there are fewer
high-energy photons) than many of the sources, the composition of the
gamma-ray sky at higher (TeV) energies changes. 
   
This review summarises the status of gamma-ray astronomy obtained
through space-based and ground-based observations. It does not aim to
provide a complete list of gamma-ray sources, nor does it aim for even
a complete list of the different source classes. It aims to illustrate
key science questions and describe their status by using selected
observational results. The interested reader is referred to recent
reviews covering related areas of research, such as a review of TeV
astronomy~\citep{HintonHofmann}, a review of pulsars at gamma-ray
energies~\citep{Caraveo}, a review of supernova remnants at high
energies~\citep{Reynolds}, reviews of dark matter observation with
gamma rays~\citep{DMReviewBergstroem, DMReviewConrad, DMReviewFunk,
  DMReviewBertone}.

\begin{textbox}
\section{The Key Science Questions}
Photon radiation in the gamma-ray band differs fundamentally from
lower energy radiation: MeV to TeV gamma rays cannot be created
thermally in celestial systems. Instead, more exotic processes allow
the concentration of large amounts of energy into a single quantum of
radiation. The key science question can be broadly classified into
four main areas: 

\subsection{The origin of Cosmic rays}
The acceleration of particles to the highest energies in astrophysical
objects is of fundamental importance for high-energy
astrophysics. Main questions that can be addressed by MeV to PeV
gamma-ray observations are: a) which sources provide the bulk of the
Galactic CRs, b) what is the spectrum of the accelerated particles and
c) what is the particle acceleration mechanism responsible for CRs?

\subsection{Dark matter annihilations}
The observation of gamma rays from the Universe provides a chance of
observing indirect (self-annihilation) signatures of DM. Numerous and
diverse hints point to the existence of weakly interacting massive
particles (WIMPs) with masses in the range $\sim 0.01-10$~TeV as the
most plausible form of DM.  If true, gamma-ray observations offer a
compelling way to connect signals found in the Large Hadron Collider
at CERN or in direct detection experiments to the actual distribution
of DM in the Universe. Gamma rays also allow for the detection of
signatures of axions via the distortion of energy spectra of distant
sources through axion-photon mixing in intervening magnetic fields.

\subsection{Relativistic Outflows}
Particle acceleration occurs in relativistic outflows such as winds or
jets. These collimated streams of relativistic plasma are commonly
seen in many prominent gamma-ray source classes, such as active
galactic nuclei, gamma-ray bursts, and in Galactic binary sources. The
accretion of material onto black holes and other very compact objects
often leads to the formation of a collimated jet of plasma that
travels outward along the axis of the accretion disk. The formation,
internal structure, and evolution of collimated jets are still not
fully understood. Gamma-ray observations can help to reveal the most
energetic of these relativistic outflows and can add to our
understanding of the processes governing these enigmatic objects.

\subsection{Cosmological questions}
Active Galactic nuclei (AGN) comprise the largest number of gamma-ray
sources detected by now. Given the large distances between us and many
of those extragalactic gamma-ray emitters, they can be used to study
properties of the intervening space. As such, gamma rays can be used
to put constraints on intergalactic magnetic and photon radiation
fields in the Universe. Additionally, gamma rays can be used to probe
voids in the Universe and test fundamental properties of quantum
gravity.\\

This review focusses on the origin of cosmic rays and the search for
dark matter annihilation. For the other two topics the reader is
referred to reviews such as ~\citep{ReviewOutflows, ReviewGammaRays}.
\end{textbox}

\section{The Science Case for gamma-ray astrophysics}

The suggestion that gamma rays traveling in straight line can reveal
the sites of proton (and electron) acceleration in the Universe still
constitutes the basis of one of the major objectives of the field.
This concerns both the acceleration and propagation aspects of the
study of CRs: while discrete gamma-ray sources indicate the sites of
particle acceleration, the spatial distribution of the diffuse
component of radiation provides information about the propagation of
relativistic particles in the ambient magnetic fields. Gamma rays are
produced in the interaction of accelerated {\emph{charged}} particles
(nuclei or electrons or positrons) with interstellar (target) material
or magnetic or radiation fields. Once these {\emph{targets}} for the
radiation, i.e. the gas densities (pion-decay and Bremsstrahlung),
magnetic (synchrotron emission) or radiation fields (inverse Compton
scattering) are known through multi-waveband studies the detection of
gamma rays allows for a determination of the detailed spatial and
energy distribution of high-energy charged particles in our
Universe. While the exploitation of this potential remains one of the
highest priority motivations of the field, the observational and
theoretical studies of recent years revealed new exciting aspects.
Most importantly, the physics and astrophysics of relativistic
outflows such as pulsar winds and active galactic nuclei (AGN) jets
and the exciting issues related to the cosmological probes of the
extragalactic radiation and magnetic fields using the beams of
gamma rays emitted by objects located at large redshifts. In addition,
the study of cosmic gamma rays promises interesting potential related
to the indirect search for Dark Matter~\citep[e.g.][]{DMLines} and the
exploration of possible deviations in fundamental rules of physics
like violation of the Lorentz invariance~\citep[e.g.][]{LIVAmelino}.

\subsection{Particle Acceleration in astrophysical objects}
\label{sec::acceleration}
A long-standing mystery is the mechanism by which CRs (protons,
electrons and positrons as well as higher nuclei) are accelerated in
our galaxy. The main properties of CRs can be derived from the locally
detected spectrum of these particles. They consist mostly of protons
with smaller fractions of higher nuclei and electrons. The
all-particle spectrum of CRs can be described by a power-law in energy
$dN/dE \sim E^{-\Gamma}$ with an index of $\Gamma = 2.7$ below the
{\emph{knee}} at $\sim 10^{15}$ eV (1 PeV) above which the index
changes to $\Gamma \sim 3.1$ continuing to about $10^{19}$ eV. It is
generally assumed that the Galactic CRs below the knee are produced by
a single mechanism (and probably single source class), whereas both
galactic and extragalactic sources have been suggested above the
knee.
The remnants of supernova explosions could sustain the flux of CRs if
they were to put $\sim 10\%$ of their bulk (explosion energy) into the
acceleration of CRS.  Whatever the sources of CRs are: to populate the
Galactic CR pool each source must on average provide of order
$10^{48}$ ergs year$^{-1}$ in protons with a power law in energy and a
spectral index of $\sim 2.2$ up to energies of $\sim 10^{15}$ (at
least for part of their lifetime). Also the acceleration of electrons
is an important topic.  Beyond the study of shock accelerating
objects, relativistic outflows, such as collimated jets from black
holes and other compact objects are important test cases for particle
acceleration in the universe

\subsection{Dark Matter Annihilation}
In the standard cosmological model, dark matter (DM) dominates the
mass density in the Universe and is the driving force in structure
formation. A favoured candidate for dark matter particles are weakly
interacting massive particles (WIMPs)\citep[see
e.g.][]{SteigmanTurnerWIMP, Feng2010}, initially in equilibrium in the
early Universe, but then frozen out due to the rapid expansion of the
Universe. The observed DM density arises naturally if masses of DM
particle in the GeV to TeV range are assumed, and annihilation cross
sections are on the weak interaction scale. DM candidates include the
lightest and stable supersymmetric particle~\citep{EllisSS}, or
Kaluza-Klein particles predicted in theories with extra
dimensions~\citep[for a review see e.g.][]{KKIdea}. DM particles are
cold -- i.e.  non-relativistic -- and they accumulate at the centers
of galaxies, but also in substructures in the halos of galaxies. WIMP
dark matter particles are expected to mutually annihilate, giving rise
to the creation of particles and of gamma rays with energies up to the
mass of the dark matter particles (assuming the velocities are so
small that annihilation is effectively almost at rest).

The search for signs of dark matter particles is a key topic in high
energy astrophysics\citep[see][for reviews and references]{Baltz2008,
  Porter2011} and especially in high energy gamma-ray astronomy. Even
if dark matter candidates such as supersymmetric WIMPs were produced
in accelerators on Earth, it will be difficult to prove that these
particles are stable over the lifetime of the Universe, and can
account for astrophysical dark matter; an astrophysical detection will
be required. 

\subsection{Other Cosmological probes}
A violation of Lorentz invariance may manifest itself as a
modification of the energy-momentum relation of photons, as predicted
in some models of quantum gravity due to the modified structure of
space-time on the Planck scale \citep[see][]{Ellis_qg}.
Corrections to the energy-momentum relation are therefore generally
predicted to scale as some power of $E/M_{\mathrm{Planck}}$, with
$M_{\mathrm{Planck}} = 1.22 \times 10^{19}$~GeV. Effects can manifest themselves
either as a energy dependence of the speed of propagation, or by
allowing certain reaction channels which normally would be
kinematically forbidden.

To search for energy-dependent variations of the speed of light, one
needs (a) a source emitting gamma rays at a well-defined moment and a
precise measurement of arrival times, with a combined precision
$\Delta t$, (b) a wide energy coverage of gamma rays $\Delta E$ and (c
) a long propagation path $D$.
Limits obtained using TeV gamma rays from AGN approach the Planck
scale \citep{magic_qg,hess_qg}, and limits for GeV gamma rays exceed
the Planck scale \citep{Fermi_qg}.

The light emitted by stars and accreting compact objects through the
history of the Universe is encoded in the intensity of the
extragalactic background light (EBL). Knowledge of the EBL is
therefore crucial to determining the nature of star formation and
galaxy evolution. After the cosmic microwave background, the EBL
produces the second-most energetic diffuse background. Understanding
of the EBL is thus also essential for understanding the full energy
balance of the Universe. Generally, direct measurements of the EBL are
limited by Galactic and other foreground emissions. However, the
imprint of the EBL on gamma rays propagating cosmological distances
can be measured~\citep{MadauPhinney1996}. A
gamma-ray photon of energy $E_\gamma$ and an EBL photon of energy
$E_{\mathrm{EBL}}$ can annihilate into an electron-positron pair. This
process becomes most effective for collisions when
$E_\gamma \times E_{\mathrm{EBL}} \leq 2 (m_ec^2)^2$, where $m_ec^2$
is the rest mass energy of the electron. This process attenuates the
spectra of gamma-ray sources above a critical gamma-ray energy of
$E_{\mathrm{crit}}(z) \approx 170 (1 + z) - 2.38$
GeV~\citep{Franceschini2008}.  
All current results indicates a rather low level of EBL in the
universe, thereby placing strong constraints on the existence of an
unseen population of stars in the early Universe or indicating that
gamma-ray photons converting to axions and back to gamma rays in
strong magnetic fields on the way from the source to us, thereby being
unavailable to the process of photon-photon pair productions for parts
of the journey~\citep{Fermi:Axions, HESS:axions}.

\section{Gamma-ray production}

\subsection{Leptonic emission}

For the gamma-ray spectra one typically starts with an accelerated
power-law spectrum of the charged particles (potentially with a
high-energy cutoff at $E_{\mathrm{max}}$ as discussed in
section~\ref{sec::acceleration}) and subsequently calculates the
losses into photons from the different processes. Each of the
different processes has certain characteristic features that can be
used to identify the underlying production mechanism once the gamma
rays are observed.

\begin{figure}[!htb]
\includegraphics[width=0.8\textwidth]{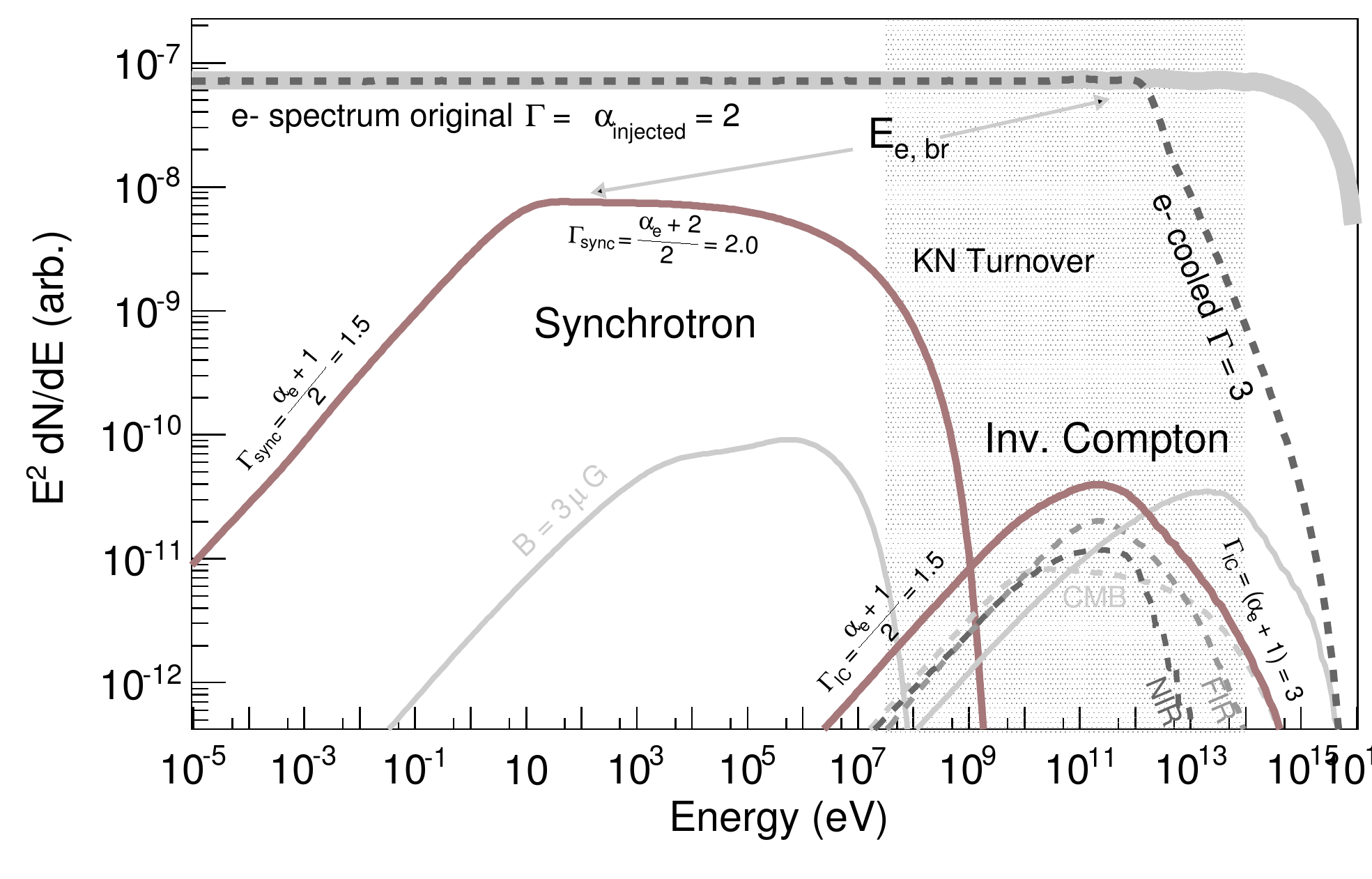}
\caption{Spectral energy distribution of electrons at injection (light
  gray with $\alpha_{\mathrm{injection}} = 2.0$) and the steady state
  including cooling (dashed dark gray) for a source with age
  $t_{\mathrm{age}} = 1000$ yrs, $B = 100 \mu G$, for a scenario in
  the inner 100 pc of our Galaxy. The cooling break in the electron
  spectrum at $\sim 1.2$ TeV is apparent in the steady-state electron
  distribution (dashed dark gray), in the synchrotron spectrum and in
  the IC spectrum. Also apparent is the turnover in the spectra at
  even higher energies due to KN cooling which incur catastrophic
  losses on the electrons. The case for a much lower B-Field of
  $3 \mu G$ is also shown in light gray. The shaded gray region shows
  the sensitive range of current gamma-ray detectors
  ({\emph{Fermi}}-LAT, IACTs).}\label{fig::spectrumElectron}
\end{figure}

The spectral shape of the synchrotron spectrum is strongly peaked with
a tail to higher energies. For an isotropic distribution of pitch
angles, a population of mono-energetic electrons with
energy $E_e$ will emit synchrotron photons at an energy $E_{\mathrm{sync}}$
with~\citep{Aharonian1997}:  
\begin{equation}
\label{eq::Esync}
  E_{\mathrm{sync}}  = 0.2 \frac{B}{10 \mu G} \left( \frac{E_e}{1\mathrm{TeV}} \right) ^2 \mathrm{eV}
\end{equation}
The synchrotron radiation spectrum of TeV electrons in a typical 10
$\mu G$ magnetic field thus peaks at approximately 0.2 eV (i.e. in the
visible range of the electromagnetic spectrum).  A more realistic case
is one where the electron population has a distribution of energies
that follows a power-law with index
$\alpha_e$~\citep{BlumenthalGould}. The differential synchrotron
spectrum in this case follows a power-law with index
$\Gamma_{\mathrm{sync}} = (\alpha_e + 1)/2$. Energy losses in the IC
Thomson regime and for synchrotron emission are proportional to
$E_e^{-1}$. These losses will therefore modify the initial power-law
distribution of electrons so that the steady state energy spectrum of
the electrons will have a break from $\alpha_{\mathrm{injected}}$ to
$\alpha_{\mathrm{injected}} + 1$ (see
Figure~\ref{fig::spectrumElectron}). The break will be at an energy
where the cooling time scales become comparable to the age
$t_{\mathrm{age}}$ of the source~\citep{Aharonian1997} and can be
approximated as
$E_{\mathrm{e, br}} = 1.2 \times 10^4 (B/ 10 \mu G)^{-2}
(t_{\mathrm{age}}/10^4 \mathrm{yr})^{-1}$
GeV. This break will induce a corresponding break in the synchrotron
and IC spectrum by $\Delta \Gamma = 0.5$ at an energy that can be
determined by inserting $E_{\mathrm{e, br}}$ into
equation~\ref{eq::Esync}.

For electrons, the inverse Compton scattering of mono-energetic
electrons on a population of target photons (e.g. a black-body
spectrum) produces a broad spectral distribution of high-energy
photons. This distribution peaks at
\begin{equation}
\label{eq::EIC}
E_{\mathrm{IC}} = 5  \times 10^{9} \frac{E_{ph}}{10^{-3} \mathrm{eV}}
\frac{E_e}{1\mathrm{TeV}}^2 \mathrm{eV} 
\end{equation}
Due to the similarity of equations ~\ref{eq::Esync} and ~\ref{eq::EIC}
the spectra for synchrotron emission and for IC scattering have the
same shape (albeit at different energies).  Like in the case of
synchrotron emission, for a continuous injection of electrons with a
power-law distribution of the form $dN/dE \propto E_e^{-\alpha}$ the
inverse Compton spectrum in the Thomson regime will have a slope of
$\Gamma = (\alpha + 1)/2$.  In the KN regime the IC spectrum will be
significantly steeper $\Gamma = (\alpha + 1)$. Therefore, even a
power-law distribution of electrons will produce a break in the
spectrum of the gamma-ray emission due to the onset of the KN regime.

\subsection{Hadronic emission}

\begin{figure}[!htb]
\includegraphics[width=0.8\textwidth]{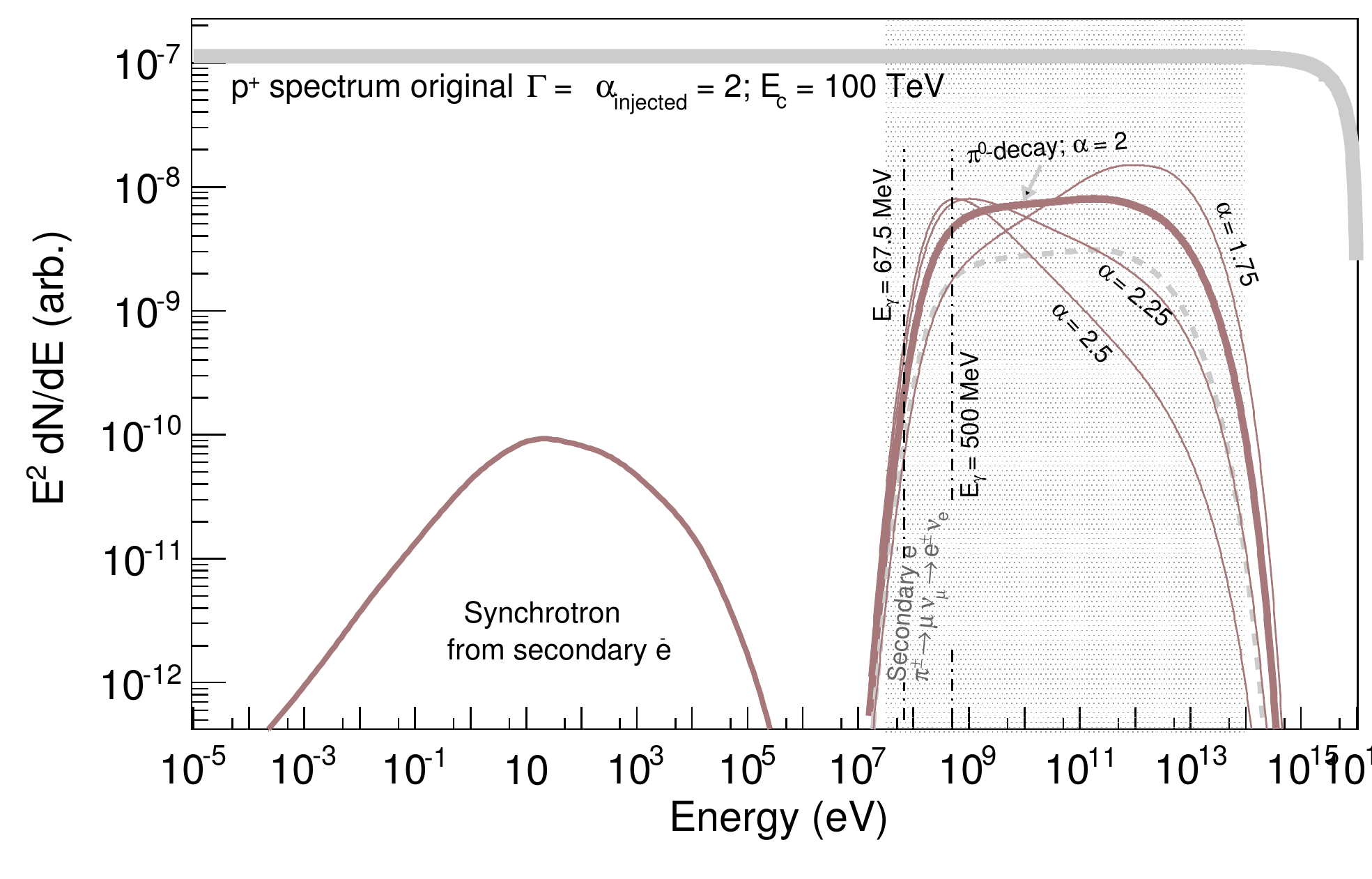}
\caption{Spectral energy distribution of accelerated protons (powerlaw
  index $\alpha_{\mathrm{injection}} = 2.0$ and cutoff at 100 TeV) and
  gamma rays resulting from inelastic collisions with interstellar
  material. The dominant emission into photons is via the decay
  $\pi^0 \rightarrow \gamma \gamma$ (solid brown). As can be seen, the
  gamma-ray spectrum follows the parent protons spectrum rather
  closely in the mid-energy range and the high-energy cutoff
  region. For all proton indices the low-energy turnover is a
  characteristic feature of the pion-decay emission. Also shown is the
  spectrum of electrons resulting from the inelastic pp-interactions
  via the decay chain
  $\pi^{\pm} \rightarrow \mu + \nu_\mu \rightarrow e^{\pm} \nu_e$
  (dashed gray).  For the synchrotron emission from these so-called
  secondary electrons a source with age $t_{\mathrm{age}} = 1000$ yrs,
  and $B = 30 \mu G$ has been assumed. The shaded gray region shows
  the sensitive range of current gamma-ray detectors
  ({\emph{Fermi}}-LAT, IACTs).}\label{fig::spectrumProton}
\end{figure}

Figure~\ref{fig::spectrumProton} shows the gamma-ray spectral energy
distribution (SED) for a proton spectrum with $\alpha = 2$,
$E_c = 100$TeV. Cooling plays a relatively minor role in sources
actively accelerating particles, since even in the case of a typical
Galactic density $n=1 \mathrm{cm}^{-3}$ the cooling time is of the
order of $10^7$ years. The shape of the gamma-ray energy spectrum away
from the threshold directly mirrors the shape of the parent proton
spectrum. 
The total fraction of the energy of each incident proton converted
into gamma rays is approximately $\kappa = 0.17$. It has been
shown~\citep[see e.g.][]{DAV1994} that for proton spectrum indices of
$2.1-2.7$ the emissivity, i.e. the number of gamma rays produced per
H-atom in the interaction of accelerated protons with interstellar
material is
$q_\gamma (>100 \mathrm{MeV}) \approx 0.5 \times 10^{-13} s^{-1}
\mathrm{erg}^{-1} \mathrm{cm}^3 (\mathrm{H-atom})^{-1} $.
This can be turned into a flux at Earth by an astrophysical
accelerator putting a fraction $\epsilon_{\mathrm{CR}}$ of its energy
output $E_{\mathrm{pr}}$ into acceleration of protons:
\begin{equation}
  F_\gamma(>100 \mathrm{MeV}) = 4.4 \times 10^{-7} \epsilon_{\mathrm{CR}} \frac{E_{pr}}{10^{51}
    \mathrm{erg}} \frac{d}{1\mathrm{kpc}}^{-2} \frac{n}{1 \mathrm{cm}^{-3}} \mathrm{cm}^{-2} s^{-1}
\end{equation} 
I.e. if distance $d$ and density of the interaction region $n$ is
known, the amount of energy in protons $E_{pr}$ at the interaction
site can be directly inferred from the gamma-ray flux $F_\gamma$.  It
should be noted, that for an energy-dependent diffusion coefficient
high-energy particles might propagate ahead of lower-energy particles.
Dense regions ahead of the shock would therefore not be fully
permeated by the particles and therefore ``see'' particles with an
effective low-energy cutoff. Particles with different energies would
therefore encounter different gas density and the resulting gamma-ray
spectrum will no more follow the proton spectrum ~\citep[see e.g.][for
a comprehensive discussion]{GabiciAharonianBlasi2007}.

Given the similarity of the gamma-ray spectra emitted by accelerated
electrons and those emitted by accelerated protons, the low-energy
part becomes a crucial tool in distinguishing the different
scenarios. For kinematic reasons, the decay of
$\pi^0 \rightarrow \gamma \gamma$ imparts each gamma ray with an
energy $E_\gamma = m_{\pi^0}c^2/2 = 67.5$ MeV in the rest frame of the
neutral pion. The resulting gamma-ray number spectrum
$dN_\gamma/dE_\gamma$, is thus symmetric about $67.5$ MeV in a log-log
representation~\citep{Stecker1970}.  The $\pi^0$-decay spectrum in the
usual $E_\gamma^2 dN_\gamma/dE_\gamma$ representation rises steeply up
to $\sim 400$ MeV (the exact turnover in this representation depends
on the parent proton spectrum as shown in
Figure~\ref{fig::spectrumProton}). This characteristic spectral
feature (often referred to as the ``pion-decay bump") uniquely
identifies $\pi^0$-decay gamma rays and thereby high-energy protons,
allowing a measurement of the source spectrum of CRs.

Electron bremsstrahlung and pp inelastic scattering both depend on the
density of ambient medium $n_0$. Assuming that electrons and protons
are accelerated with the same power-law spectrum, the ratio of
gamma-ray emissivities (i.e. the emission rate per hydrogen atom) can
be estimated as
$q_\gamma^{br} / q_\gamma^{\pi^0} \sim R 3 \tau_{pp}/\tau_{Br} = 4 R$
where $R$ is the electron to proton ratio. For typical values of
$R \leq 10$ as in the expected sources of Galactic CRs pion production
at high energies dominates over bremsstrahlung gamma rays.

\subsection{Dark Matter Decay or Annihilation}

\begin{figure}[!htb]
\includegraphics[width=0.8\textwidth]{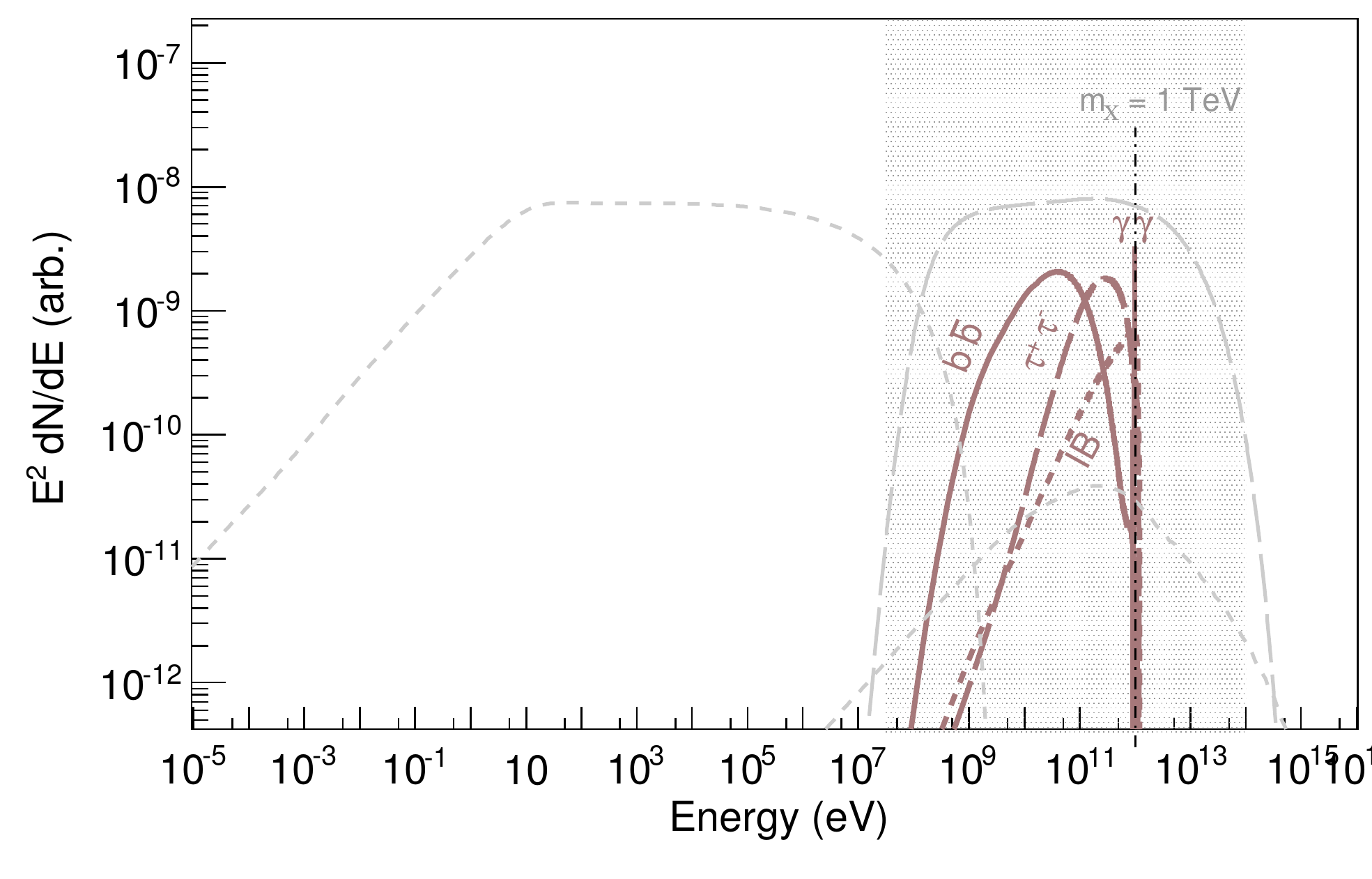}
\caption{Spectral energy distribution of gamma rays resulting from
  neutralino annihilation solid brown) with mass $m_{\chi} = 1$ TeV.
  The spectrum consists of 4 (main) components: continuum emission
  from quarks and bosons (labeled $b \bar{b}$ shown here as an
  example), continuum emission from heavy leptons (labeled
  $\tau^+ \tau^-$), internal bremsstrahlung from annihilation into
  leptons ($e^+e^-$ and $\mu^+ \mu^-$ labeled IB) and line emission
  into $\gamma \gamma$ and $Z\gamma$ etc).  The relative strength of
  these components depends on the particle physics model chosen for
  the annihilation. The typical astrophysical foreground spectra from
  accelerated electrons and accelerated protons are also shown in
  light gray (see Figures~\ref{fig::spectrumProton} and
  \ref{fig::spectrumElectron}) for comparison.  The shaded gray region
  shows the sensitive range of current gamma-ray detectors
  ({\emph{Fermi}}-LAT, IACTs).}\label{fig::spectrumAnnihilation}
\end{figure}

The flux of gamma rays from annihilation processes is given by a
product of a factor depending on particle physics properties of DM
particles, and a factor depending on their astrophysical distribution
\citep{Bergstrom1998}
\begin{equation}
 \label{eq::annihilation}
\phi = \left({<\sigma v> \over M^2} {dN_\gamma \over
    dE}\right)\left({1 \over 4\pi} \int_{\mathrm{LOS}}\rho^2dl \right)
\end{equation}
where $<\sigma v>$ is the velocity-weighted annihilation cross section
of DM particles of mass $M$, $dN_\gamma/dE$ the gamma-ray spectrum per
annihilation event, and the second factor is the line-of-sight
integral of the squared DM particle density $\rho^2$. The gamma-ray
emission from DM annihilation can often be described as a combination
of several processes and depends strongly on the annihilation channel
(see Figure~\ref{fig::spectrumAnnihilation}). The most common of these
contribution is usually the hadronisation of decay or unstable
particles. Supersymmetric models typically predict the annihilation of
the lightest supersymmetric particle (often the neutralino) into heavy
final states consisting of $b \bar{b}$, $t \bar{t}$ or $\tau^+ \tau^-$
or $ZZ$, $W^+ W^-$ etc. Each of these channels (with the exception of
annihilation into $\tau^+ \tau^-$) produce very similar
({\emph{continuum}}) spectra of gamma rays, ultimately dominated by
the decay of (secondary) mesons such as $\pi^0$ and therefore
resembling the broad spectra seen in the more traditional (IC or
$\pi^0$ decay) channels discussed above with a cutoff at the dark
matter mass (see Figure~\ref{fig::spectrumAnnihilation} labeled
$b\bar{b}$). The spectra are softer for annihilation into $b$ quark
pairs, compared to light quark pairs. Annihilation may also proceed
directly into the two-gamma (or the $\gamma Z$) channel, via a
loop-level process, resulting in gamma-ray lines~\citep{DMLines} as
perfect signature, but the cross sections are in most models strongly
suppressed. If helicity arguments disfavor annihilation into two
fermions, as is the case in certain classes of models, radiative
corrections \citep[so-called internal
Bremsstrahlung][]{DMInternalBrems} may enhance annihilation channels
where the gamma-ray spectrum mimicks a broad line near the kinematic
limit (see Figure~\ref{fig::spectrumAnnihilation} labeled IB). For a
more detailed review see e.g.\citep{DMReviewBergstroem,
  DMReviewConrad}.

\section{Status of instrumentation}
Gamma rays span an energy range of about 7 decades in energy and
subsequently a typical flux difference of 14 decades between the low
and the high-energy end. Consequently, not one technique or instrument
can cover the whole energy range. While at the low energy end a bright
source such as e.g. the Vela pulsar exhibit $\approx 10^{-1}$ photons
m$^{-2}$ s$^{-1}$ (or the Crab Nebula $\approx 10^{-3}$ photons
m$^{-2}$ s$^{1}$) above 100 MeV, at the high energy end the brightest
source such as the Crab Nebula exhibit $\approx 10^{-7}$ photons
m$^{-2}$ s$^{1}$. While one usually wants to build instruments with
detection areas as large as possible, at low energies
$\mathcal{O}(1 m)$ are appropriate, while at high-energies areas
($> 50$ GeV) $\mathcal{O}(10^4m)$ are needed.

\begin{figure}[!htb]
\includegraphics[width=\textwidth]{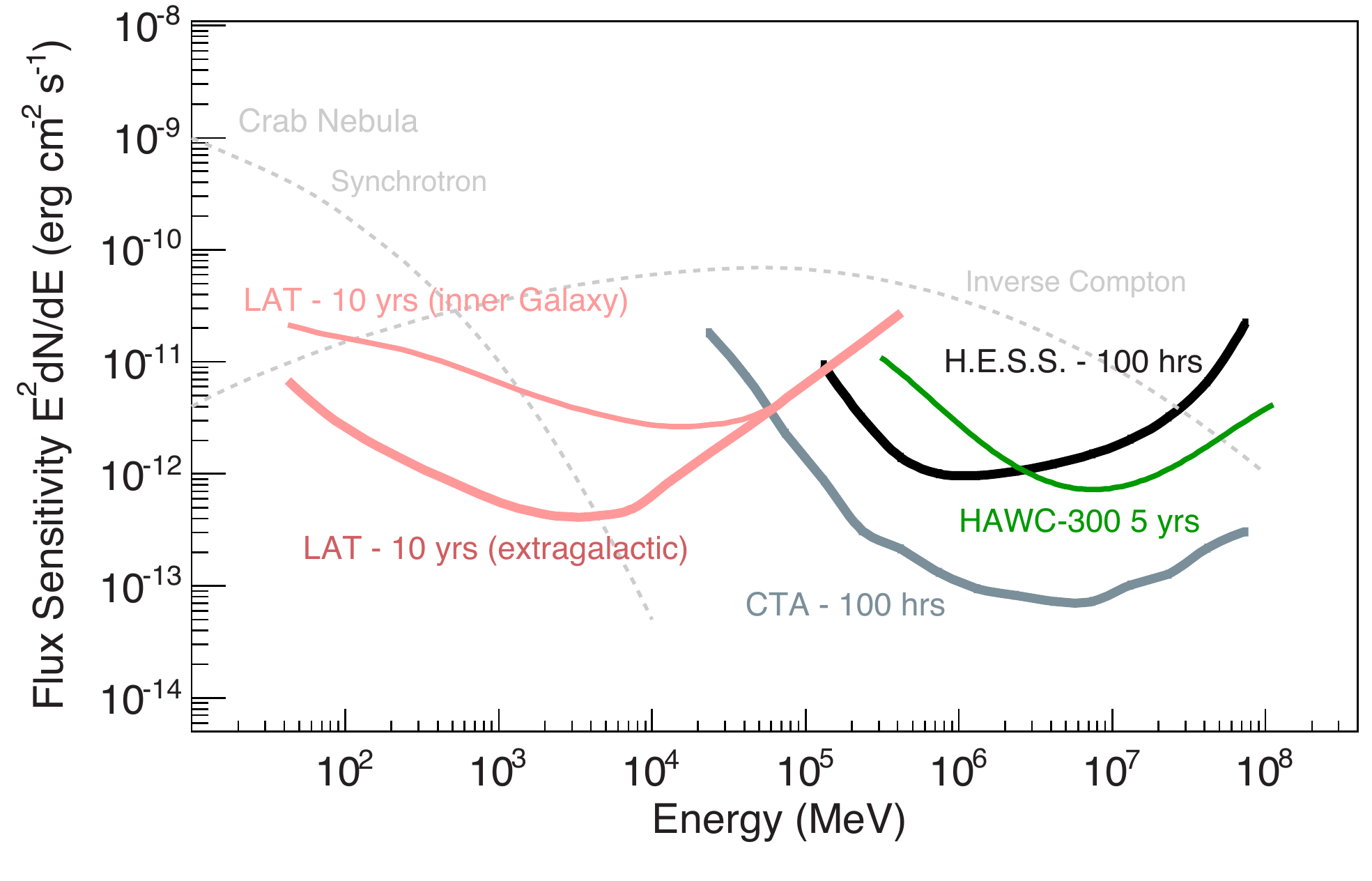}
\caption{``Differential'' sensitivity (integral sensitivity for 4 bins
  of energy per decade) for a minimum significance of $5\sigma$ in
  each bin, minimum 10 events per bin. For {\emph{Fermi}}-LAT, the
  curve labeled ``inner Galaxy'' corresponds to the background
  estimated at a position of $l = 10^{\circ}, b = 0^{\circ}$, while
  the curve labeled ``extragalactic'' is calculated for
  $l = 10^{\circ}, b = 90^{\circ}$. For the ground-based instruments a
  5\% systematic error on the background estimate has been
  assumed. For the {\emph{Fermi}}-LAT, {\emph{Pass8SourceV5}}
  instrument response function curves have been used (number of events
  n = 10 and TS = 25 per bin). For comparison, the synchrotron and
  Inverse Compton measurements for the brightest persistent TeV
  source, the Crab Nebula are shown as dashed grey curves. The HAWC
  sensitivity (in green) was reproduced from~\citep{HAWCSensitivity}.}
\label{fig::Sensitivity}
\end{figure}

However, the atmosphere blocks gamma rays from reaching the
ground. Therefore their direct detection needs space-based
instruments. At higher energies the detection area of space-based
detectors is insufficient, so that a ground-based technique is
needed. Ground-based instruments operate on the principle of detecting
the secondary products of the interaction of gamma rays with the
atmosphere. In this interaction a particle shower in the atmosphere
gets created consisting mostly of electrons and positrons (in the case
of a gamma-ray or electron-initiated showers) or of electrons and
muons (in the case of proton-initiated showers). The two main
techniques that utilize the air shower techniques are {\emph{Imaging
    Atmospheric Cherenkov Telescopes (IACTs)}} that detect the
(optical) Cherenkov light produced by the shower particles in the
atmosphere and {\emph{Water Cherenkov Detectors}} that directly detect
the particles through the Cherenkov light produced in water detectors
on the ground. Since the latter require the particles to reach the
ground, their threshold is usually higher ($\sim 100$GeV), but they
can operate continuously while the IACTs can only operate during dark
night time (to be able to detect the faint Cherenkov light from the
shower).

All detection techniques must deal with the common issue that in order
to detect a significant signal of gamma rays the much more numerous
background of CR protons has to be suppressed. This is usually done
by an active veto for charged particles (in the case of space-based
detectors) or by making use of the difference in shower development
(mostly shower shape and muon content) between a gamma-ray induced
electromagnetic shower and a proton-induced hadronic shower. Typical
suppression factors are $\sim 10^{-6}$ for the {\emph{Fermi}}-LAT, up
to $\sim 10^{-5}$ for IACTs (with an angular cut optimised for
point-sources) and several $10^{-3}$ for Water Cherenkov detectors
such as HAWC~\citep{HAWCSensitivity}.

The detection techniques are complementary in various aspects as
illustrated in Figure~\ref{fig::Sensitivity}: while the space-based
detectors are ideally suited for the low-energy part of the spectrum
starting at $\sim 30$MeV even the largest operating instruments such
as the {\emph{Fermi}}-LAT typically run out of sufficient detection area at
around $\sim 100$ GeV (for the brighter sources). This is the regime
that corresponds to the threshold energy of the IACTs where enough
Cherenkov light in the particle shower in the atmosphere gets produced
to trigger a typical $\sim 10$m mirror diameter IACT. At the highest
energies ($\sim 30$ TeV for the brighter sources) the IACT again run
out of detection area and this is the regime where covering the ground
with a large number of water Cherenkov telescopes can provide the
necessary detection areas. 
A detailed comparison of the sensitivities of space-based and future
ground-based instruments reveals that IACTs will be more sensitive
than the {\emph{Fermi}}-LAT for measuring spectra of sources with an
$E^{-2}$ spectrum above $\sim 40$GeV~\citep{FunkHinton}.

Given all these different observational efforts the current time can
be seen as ``Golden age" for gamma-ray astronomy with a variety of
currently operating instruments that if used in conjunction offer
observations over more than 7 decades in energy with high angular
resolution (typically up to $\sim 0.1^{\circ}$) and significant
coverage of the sky at any given moment.

\subsection{Space-based observatories}

Space-based detectors above 20 MeV operate on the principle of
pair-creation in the detector. The main instrument in this energy
range is the {\it Fermi Gamma-ray Space Telescope} (FGST, formerly
GLAST) launched in June 2008. The {\it Fermi} LAT (Large Area
Telescope) yields observations in the energy range between $\sim 20$
MeV and $\sim 300$ GeV. The instrument consists of a {\emph{tracker}}
to measure the tracks of the electron-positron pair created in the
pair-creation of the gamma ray, a {\emph{calorimeter}} to determine
the energy of this electron-positron pair (and hence of the primary
gamma ray) and an {\emph{anti-coincidence detector}} to veto the
charged particle background. The tracking section of the LAT has 36
layers of silicon strip detectors interleaved with 16 layers of
tungsten foil to facilitate the pair-creation (12 thin layers, 0.03
radiation length, at the top followed by 4 thick layers, 0.18
radiation length, in the back section, adding up to a total radiation
length of $\sim 1 X_0$). The total surface area of the LAT is
$\approx 1.8\mathrm{m} \times 1.8\mathrm{m}$ with about 80\% of that
being sensitive tracker area. The calorimeter is composed of an
8-layer array of CsI crystals (∼8.5 total radiation lengths). The
segmented anti-coincidence detector surrounding the tracker consists
of plastic scintillators read out by photomultiplier tubes. The
instrument typically triggers at a rate of $\sim 3$kHz, completely
dominated by the proton background. About 90\% of the events are
discarded on the fly resulting in a down-link rate of about 450 Hz
(90\% of these are events that pass the gamma-ray filter, with an
additional 25 Hz of diagnostic filter and 5 Hz of a Heavy ion
filter). On the ground, these events are reconstructed and additional
gamma-hadron seperation cuts applied. The total photon rate after
background rejection cuts is a few Hz.

The effective collection area of the LAT (taking into account gamma
rays that are lost during the removal of the charge particle
background) is about $0.65 \mathrm{m}^2$ above $\sim 1$GeV.  For this
energy range, the {\emph{Fermi}}-LAT provides an unprecedented angular
resolution ($0.8^\circ$ at 1 GeV averaged over the LAT acceptance
and better than $0.2^\circ$ at energies above 10 GeV).  It is a well
designed tool for deep surveys of the gamma-ray sky with a large field
of view of 2.4 steradian (at 1 GeV) \citep{FermiLATInFlight}. The
live time of the instrument is $\approx 75\%$ -- the main factors
limiting this fraction being the South Atlantic Anomaly~\citep{SAA}
when the instrument is switched off (13\%) and the readout dead-time
fraction (9\%). More information about the design of the LAT is
provided
in~\citep{LATPaper}, the in-flight calibration of the LAT is described
in \citep{FermiLATInFlight}. One advantage over ground-based detectors
is that the instruments can be calibrated in the laboratory (since the
atmosphere is not part of the detector).  {\emph{Fermi}}-LAT has been tested in
various beam tests~\citep{Fermi:BeamTest} before launch. The resulting
systematic error in the effective area of the {\emph{Fermi}}–LAT has been
estimated to be 10\% below 100 MeV, 5\% at 562 MeV, and 20\% above 10
GeV with linear interpolation in logarithm of energy between the
values~\citep{Fermi:egb}.

Over the last 7 years, {\it Fermi} LAT has provided a significant
improvement in our understanding of the MeV to GeV gamma-ray sky (see
e.g.\ Fig.\ref{fig::FermiSky}). The results confirm the (optimistic)
pre-launch expectations that included a dramatic increase of the
number of gamma-ray emitting pulsars and AGN (more than 1200 by now),
discovery of new classes of {\it compact/variable} and {\it extended}
galactic and extragalactic gamma-ray sources,

Important results have also been contributed by the Italian gamma-ray
satellite AGILE (Astrorivelatore Gamma ad Immagini LEggero - Light
Imager for Gamma-ray Astrophysics)\citep{AGILE}. AGILE's main
instruments, the gamma-ray imager (GRID) is similar to {\emph{Fermi}}-LAT but
on a smaller scale. It consists of a Tungsten-Silicon tracker and of a
CsI calorimeter. The tracker consists of 14 planes of area
$\sim 40\mathrm{cm} \times 40 \mathrm{cm}$ (i.e. significantly smaller
surface area than the LAT) with a total radiation length of $1 X_0$
(like in the case of the LAT). The on-axis effective area is about
0.04 $\mathrm{m}^2$ (compare to $\sim 0.8 \mathrm{m}^2$ for the LAT).

\subsection{Ground-based observatories}
Ground-based observatories have in common that they detect the
secondary products of the interaction between the gamma ray and the
atmosphere (the shower particles). While IACT detect the Cherenkov
light of the shower particles in the atmosphere, Water Cherenkov
detectors detect the Cherenkov light of the shower particles in the
detector (the water tanks). These detectors typically have three
operating regimes: at the low energy end (the threshold region) the
sensitivity is governed by the ability to trigger the detector and the
relevant question is whether the energy of the gamma ray is large
enough to produce enough Cherenkov light in the atmosphere (or
energetic enough showers to reach the ground) to trigger the detector.
The thickness of the atmosphere at ground-level is
$\sim 1000 \mathrm{g cm}^{-2}$ which corresponds to about 1 m of lead or
$\sim 28$ radiation lengths. The shower typically starts with
pair-creation of the gamma ray into an electron positron pair at an
altitude of $\sim 20 \mathrm{km}$ above ground. The depth of the
shower maximum $X_{\mathrm{max}}$ depends logarithmically on the
energy of the gamma ray. For a 1 TeV gamma rays, $X_{\mathrm{max}}$
corresponds to an altitude of 10~km above sea level. The background of
charged particles (mostly protons) also producing air showers is up to
$10^4$ larger than rate of gamma rays. To efficiently suppress this
background the shape and composition of the air shower in the
atmosphere as well as the expected direction from which the gamma rays
are arriving are the main handles.

\subsubsection{Imaging Atmospheric Cherenkov Telescopes}

In recent years these instruments have emerged as the most sensitive
detectors for observing gamma rays above $\sim 50$ GeV. They map the
Cherenkov light of the air showers with mirrors onto a fast camera in
the focal plane of the mirrors and thereby measure the angular
distribution of the Cherenkov light from the air shower. Gamma-ray
induced showers have typical rms widths and rms length of
$0.1^{\circ}$ and $0.3^{\circ}$. A pixelation in the camera much finer
than this rms width does not seem to be very effective.

A few basic instrument requirements can be derived from the
aforementioned 1 TeV shower: the (mainly blue) Cherenkov light arrives
in a ring on the ground with a radius of $\sim 120$m at an altitude of
2000 m. Stereoscopic observations have been demonstrated to be
important for background rejection (significantly suppress hadronic
showers in which long-lived muons can trigger individual telescopes at
the trigger level) and for angular resolution. The spacing of the
telescopes should thus be $\sim 100$m. Closer spacing allows for an
improvement of the low-energy performance at the expense of collection
area at higher energies (and vice versa). The total amount of light is
proportional to the total track length of all the particles in the
shower -- which in turn is proportional to the energy of the primary
particle -- and thus provides a calorimetric measurement of the
shower. Typically, $\sim 100$ photons $m^{-2}$ per TeV of primary
energy reach the ground at 2000 m altitude; at sea level this number
drops to $\sim 10 m^{-2}$. For typical efficiencies of IACTs to detect
photons ($\sim 10\%$ -- mirror losses, quantum efficiency of
detectors) $\sim 100 m^2$ of mirror area are needed to trigger a
telescope with 100 photo-electrons for a 0.1 TeV shower.  The
thickness of the light front is $\sim 1$m -- corresponding to an
arrival time of a few nanoseconds.  The night sky background at dark
sites is typically $\sim 2 \times 10^{12}$ photons
$(m^2 \mathrm{s} \mathrm{sr})^{-1}$ (about 9 orders of magnitude
smaller than the day-time sky background), so that for a typical
pixels in the camera with field of view of diameter $0.2^{\circ}$ and
an effective mirror area of $\sim 60 m^2$ like in H.E.S.S. (after
accounting for shadowing, mirror reflectivity, Winston cone losses
etc.), the night sky background is $\sim 1 \times 10^{9}$ photons
$s^{-1}$, corresponding to $\sim 100$MHz of photo-electron rate.
For these reasons IACTs have to observe during dark night times and
with extremely short exposure times of a few times $10^{-9} s$ to be
able to detect the faint Cherenkov light from the shower over the
night-sky background in the camera pixels. The duty cycle of these
instruments is thus limited to dark nights which amounts to
$\sim 10\%$. Three currently operating observatories use the
stereoscopic IACT technique: H.E.S.S., MAGIC and VERITAS.

 H.E.S.S. ({\emph{High Energy Stereoscopic System}}) is a system of
originally four (now five) IACTs located in the Khomas highland (at
an altitude of 1800 m) in Namibia in southern Africa. The original
four H.E.S.S.\ telescopes have a mirror area of 107 $m^2$ with a focal
length of 15m. The energy threshold at zenith is $\sim 100$ GeV (see
Figure~\ref{fig::Sensitivity}) at the trigger level. 
In Phase-II of the project a single very large dish with
$\sim 600 m^2$ mirror area and a focal length of 36$m$ was added to
the center of the array to improve the low-energy performance and to
lower the energy threshold to $\sim 30$ GeV, significantly increasing
the overlap with the {\emph{Fermi}}-LAT range~\citep{HESS:phaseII}.

The MAGIC ({\emph{Major Atmospheric Gamma-ray Imaging Cherenkov
    Telescopes}}) telescope system consists of two telescopes located
on the Canary Island of La Palma at an altitude of $\sim 2400m$. The
two telescopes have a very large mirror area of $\sim 240 m^2$,
corresponding to an energy threshold of $\sim 50$ GeV (even down to
$\sim 25$ GeV for special operations in so-called sum-trigger
mode). They are built in a light-weight fashion to allow for fast
slewing to be able to rapidly follow up on the alert of gamma-ray
bursts~\citep{MAGICII:status}

VERITAS ({\emph{Very Energetic Radiation Imaging Telescope Array
    System}}) is a system similar in sensitivity and performance to the
H.E.S.S. phase I project, i.e.\ consisting of 4 IACTs with mirror size 106
m$^2$. The camera has a somewhat smaller field of view
($\sim 3.5^{\circ}$) than H.E.S.S.-I, but has the advantage of a very
flexible trigger scheme using 500 MHz
Flash-ADCs~\citep{VERITAS:status}.

\subsubsection{Water Cherenkov Detectors}
Water Cherenkov detectors operate at a higher energy threshold than
IACTs given that they require the shower particles to reach the
ground. For that reason these detectors are typically located at
higher altitudes. The total energy reaching the ground in the form of
electromagnetic particles is roughly 10\% of the primary energy (at
5200 m above sea level).  Water Cherenkov detectors and IACTs have
complementary capabilities that allow for a deeper study of the
gamma-ray sky at TeV energies. Water Cherenkov detectors are typically
less sensitive to point sources than IACTs (due to a combination of
higher energy threshold and larger angular resolution than
IACTs). However, they continuously monitor the entire sky above the
detector. The solid angle surveyed by a Water Cherenkov detector is
$\Omega = 4 \pi cos(l) sin(\Theta_{max})$ where $l$ is the latitude of
the detector and $\Theta_{max}$ is the maximum zenith angle of the
observation (typically $45^{\circ}$). For the Sierra Negra, Mexico,
the site of HAWC ({\emph{High Altitude Water Cherenkov Gamma Ray
    Observatory}}), that results in $\Omega = 2.6 \pi$sr. The
combination of a large field of view and nearly 100\% duty cycle make
them well suited to study transient sources and perform unbiased sky
surveys. Also, for largely extended sources $> 1^{\circ}$ Water
Cherenkov detectors can often be more sensitive due to the fact that
IACTs typically determine the residual hadronic background from within
the field of view.

Water Cherenkov detectors operate on the principle of detecting the
Cherenkov light from the shower particles in the water tanks. The
detector must act as an effective calorimeter to be able to suppress
the hadronic background. Therefore, the PMTs need to be deep in the
water, to block particles passing close to the photo-cathode of the
detector and produce large pulses (that would not be proportional to
the deposited energy). 

The HAWC detector consists of 300 steel tanks each 4 meters high and
7.3 meters in diameter holding 188,000 litres of water. The site of
the observatory is the Sierra Negra, Mexico at an altitude of
4100$m$. Each of the tanks contains a watertight bladder and four
8-inch PMTs, sensitive at ultraviolet wavelengths. Three of the PMTs
are placed on the bottom of the tank looking upward, and spaced
several meters from the center of the tank. The fourth (10-inch
high-quantum efficiency) PMT is positioned in the bottom center of the
tank and is intended to improve the low-energy performance. The single
PMT hit rate for HAWC is $\sim 20$kHz and the total non-correlated hit
rate for the entire detector $\sim 20$MHz. Triggering is done on
showers that produce $> 30$ PMTs hit within a $\sim 50$ns window. The
trigger rate in this regime corresponds is $5-10$kHz. The background
rejection (at a fixed gamma-ray efficiency of $50\%$ is a function of
energy and improves from $10\%$ at 500~GeV to a few times $10^{-3}$ at
above 10~TeV.

\begin{marginnote}
\entry{FGST}{{\emph{Fermi}} Gamma-ray Space Telescope}
\entry{Fermi-LAT}{{\emph{Fermi}} Large Area Telescope}
\entry{H.E.S.S.}{High Energy Stereoscopic System}
\entry{MAGIC}{Major Atmospheric Gamma-Ray Imaging Cherenkov
  Telescopes}
\entry{VERITAS}{Very Energetic Radiation Imaging Telescope Array
  System}
\entry{HAWC}{High-Altitude Water Cherenkov Observatory} 
\entry{CTA}{Cherenkov Telescope Array}
\end{marginnote}

\section{Science Discoveries}
Our understanding of the gamma-ray sky has evolved and improved over
the last decade. The combination of instruments such as the
{\emph{Fermi}}-LAT and AGILE at GeV energies and ground-based
observatories IACTs such as H.E.S.S., MAGIC and VERITAS coupled with
Water Cherenkov detectors such as MILAGRO and now HAWC have ushered an
era of both significant increase in the number of sources as well as a
significantly improved understanding of the most prominent members of
the various source classes. In general, all sources of GeV and TeV
gamma rays are either locations of high-energy (non-thermal) processes
or are indication of locations of new physics in the Universe beyond
the standard model of particle physics.

The image of the GeV sky (Figure~\ref{fig::FermiSky}) illustrates
three main components of the gamma-ray sky: a) Galactic diffuse
emission, generated by the interaction of the pool of CRs in our
Galaxy with interstellar gas and radiation fields, b) individual
sources, and c) a faint glow of isotropic diffuse emission, detectable
at all Galactic latitudes. The relative contribution of these sources
to the total gamma-ray flux changes with increasing energies. While at
GeV energies about 80\% of all photons are Galactic diffuse emission,
at TeV energies individual sources dominate since the source spectra
are typically harder (meaning more high-energy photons) than the
Galactic diffuse emission. The latter follows the spectrum of the pool
of CRs which exhibits a power-law in energy with a photon index of
$\Gamma=2.7$ and therefore dominates mostly at low energies. The
Galactic diffuse emission is, however, also detected at TeV energies
by H.E.S.S.\ in the innermost part of the Galaxy (see discussion
below) and also by MILAGRO in a large-scale region on the Galactic
plane. Also recently, the deep H.E.S.S.\ observations of the Galactic
plane survey~\citep{HESS:scanpaper1, HESS:scanpaper2} allowed for the
first time to detect a large-scale gamma-ray emission along the
Galactic Plane using IACTs. This and the MILAGRO detection at TeV
energies could thus either have a CR-induced origin or could be due to
unresolved source populations or a combination thereof. The Galactic
diffuse emission generally provides very important information about
the propagation and diffusion of CRs and additionally on interstellar
radiation and gas fields.

\begin{figure}
\centering
\includegraphics[width=\textwidth]{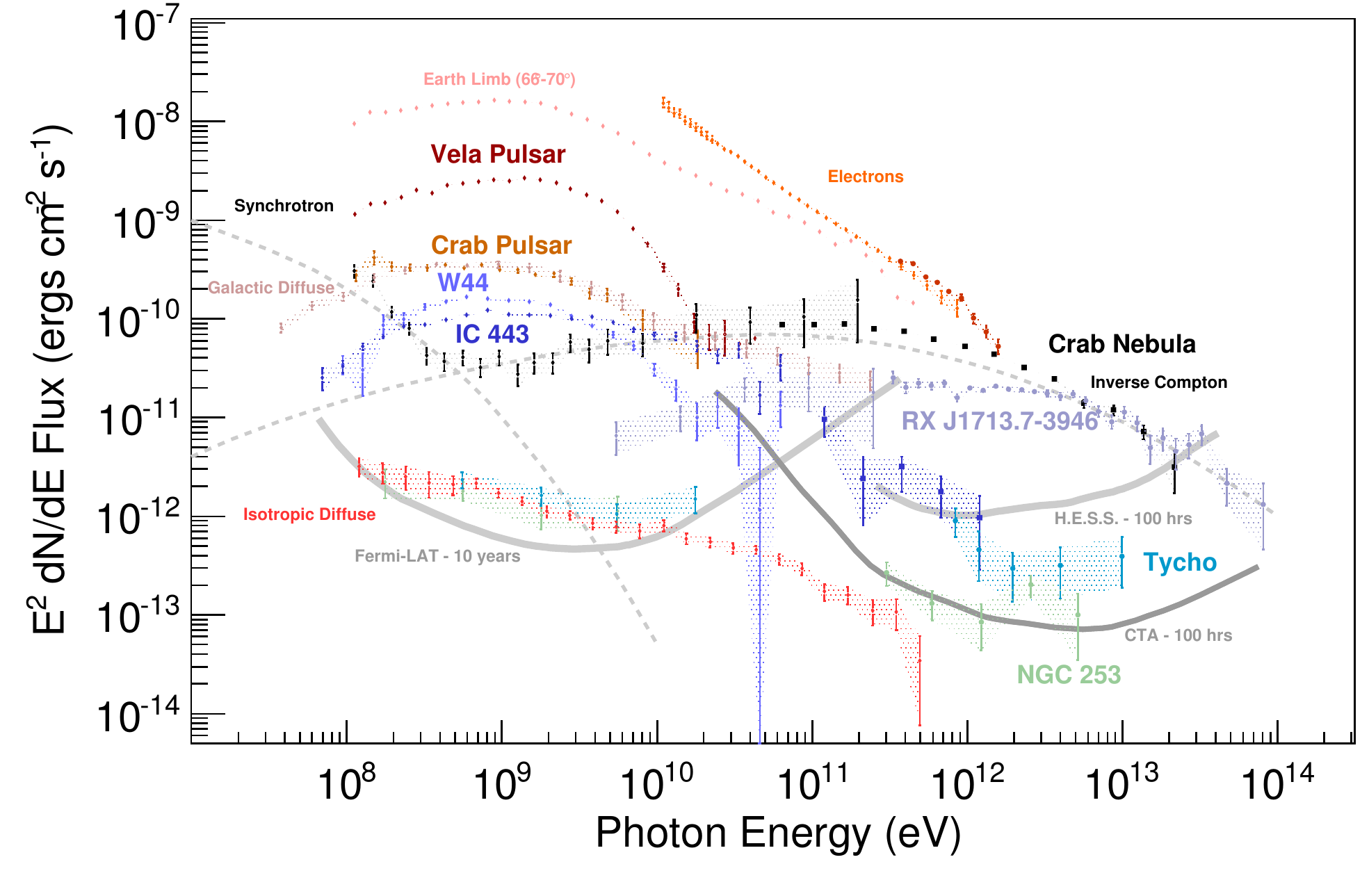}
\caption{Spectra for a selected number of GeV and TeV sources -- both
  for point-like and diffuse sources (for the diffuse sources the
  solid-angle has been converted to a field of view of a typical
  ground-based instrument of $5^{\circ}$). Also shown are the
  {\emph{Fermi}}-LAT, H.E.S.S., and CTA differential sensitivity for 4
  bins per decade. }
\label{fig::dynamicRange}       
\end{figure}

One of the most important discoveries in recent years both at GeV and
TeV energies was the variety of source classes with which the
gamma-ray sky is populated. Quite generally, given that most energy
spectra are powerlaws that decrease rapidly with increasing energy the
expectation is that for instruments with similar flux sensitivities
(as is the case with current ground and space-based instruments) the
numbers of sources at low energies will exceed the number of sources
at high energies. With ground-based instruments about $148$ sources
have been discovered both galactic and extragalactic in origin (see
e.g. Figure~\ref{fig::FermiSky}) above $\sim 100$GeV.  At GeV energies
the third {\it Fermi} LAT gamma-ray source catalog \citep{3FGL} (based
on the first four years of observations) lists more than 3000 galactic
and extragalactic gamma-ray emitters (see
Figure~\ref{fig::FermiSky}). Out of the 148 TeV sources a large
fraction (117) are also detected at GeV energies by
{\emph{Fermi}}-LAT. On the extragalactic sky all but one of the 58 TeV
AGN are detected with {\emph{Fermi}}-LAT (only HESS\,J1943+213 does
not have a 3FGL counterpart). This TeV blazar is unique in that it is
located on the Galactic plane and therefore in a region of reduced
sensitivity for {\emph{Fermi}}-LAT.

For both the GeV and the TeV sky, the identification of gamma-ray
emitting regions with astrophysical counterparts is an important
topic. Quantitatively, the 3rd {\emph{Fermi}}-LAT catalog lists about
two-thirds of the detected objects as positionally coincident
(i.e. {\emph{associated}}) with counterparts representing known source
populations (more than two hundred thirty sources firmly identified
based on the angular extent or the temporal behaviour), the origin of
approximately one-third of these objects remains an open issue. The
same is true for the TeV sky.  The limited angular resolution and (at
least on the GeV sky) the bright diffuse emission from the pool of
Galactic CRs makes source confusion in the Galactic plane a serious
issue. The most reliable approach for the identification of gamma-ray
sources is the analysis based on temporal studies. In this regard, the
best ``astronomical clocks" - the pulsars - constitute the largest
population of identified Galactic gamma-ray sources (more than 120) in
the GeV band. The spectra of pulsars do however typically show at
break in the multi-GeV region so that a detection at TeV energies
becomes extremely difficult and only possible for the brightest
objects, such as the Crab pulsar and the Vela pulsar. At TeV energies,
the largest number of detected Galactic sources are pulsar wind
nebulae (PWN) that ultimately also convert the pulsar rotational
energy into electromagnetic radiation. The periodic character of
gamma-ray emission of the Galactic binary systems or the sporadic
flares of AGN provide another tool for identification of variable
gamma-ray sources based on simultaneous observations in different
energy bands. Given the very large number of unidentified sources both
at GeV and TeV energies, interesting (and unexpected) physics might be
waiting to be discovered.

Instead of giving a laundry list of all the sources discovered and the
important science topics covered in their interpretation, in the
following, I will highlight two of the most important science cases
for gamma-ray astronomy: the origin of CRs and the search for
dark matter.

\subsection{The Origin of Cosmic rays}
All gamma-ray sources are potential sources of CR protons. Gamma rays
therefore can be used to trace the populations of energetic particles
in SNRs. The flux of hadronic gamma rays is governed by the CR density
and the target gas density. The flux of leptonic gamma rays is traced
by the electron density and the radiation fields (which are usually
assumed to be constant on the scale of the source). To distinguish the
sources of CR protons from the sources of CR electrons, gamma ray
observations are often not sufficient but multi-wavelength
observations have to be taken into account. These observations --
mostly at radio and X-ray energies -- indicate that for the best
candidate sources for the origin of Galactic CRs -- SNRs --
ultra-relativistic electrons and large magnetic fields (beyond
$100~\mu G$ for several of the young SNRs) are present in the
shocks. If indeed large magnetic fields are present in SNR shocks, the
two conclusions that can be drawn are: a) the gamma-ray emission is
probably hadronic in origin, since the electron density needed to
explain the synchrotron flux is rather low, b) the best-understood way
to enhance or amplify the magnetic field in SNR shock front is though
the pressure of accelerated protons. This so-called streaming
instability of upstream CRs in a parallel shock~\citep{BellLucek,
  LucekBell2000} is a matter of active research and its existence has
strong implications for the maximum energy achievable in SNR
shocks. Young SNRs that show indication of large magnetic fields are
ideal targets to search for gamma-ray emission that is hadronic in
origin and for sources of CRs up to very high energies -- possibly
even close to the knee in the spectrum of CRs at $10^{15}$ eV. SNRs
for which the shock wave is encountering a region of dense
interstellar material such as a molecular cloud can be expected to
have a high flux of hadronic gamma rays.

\begin{figure}[!htb]
\includegraphics[width=0.8\textwidth]{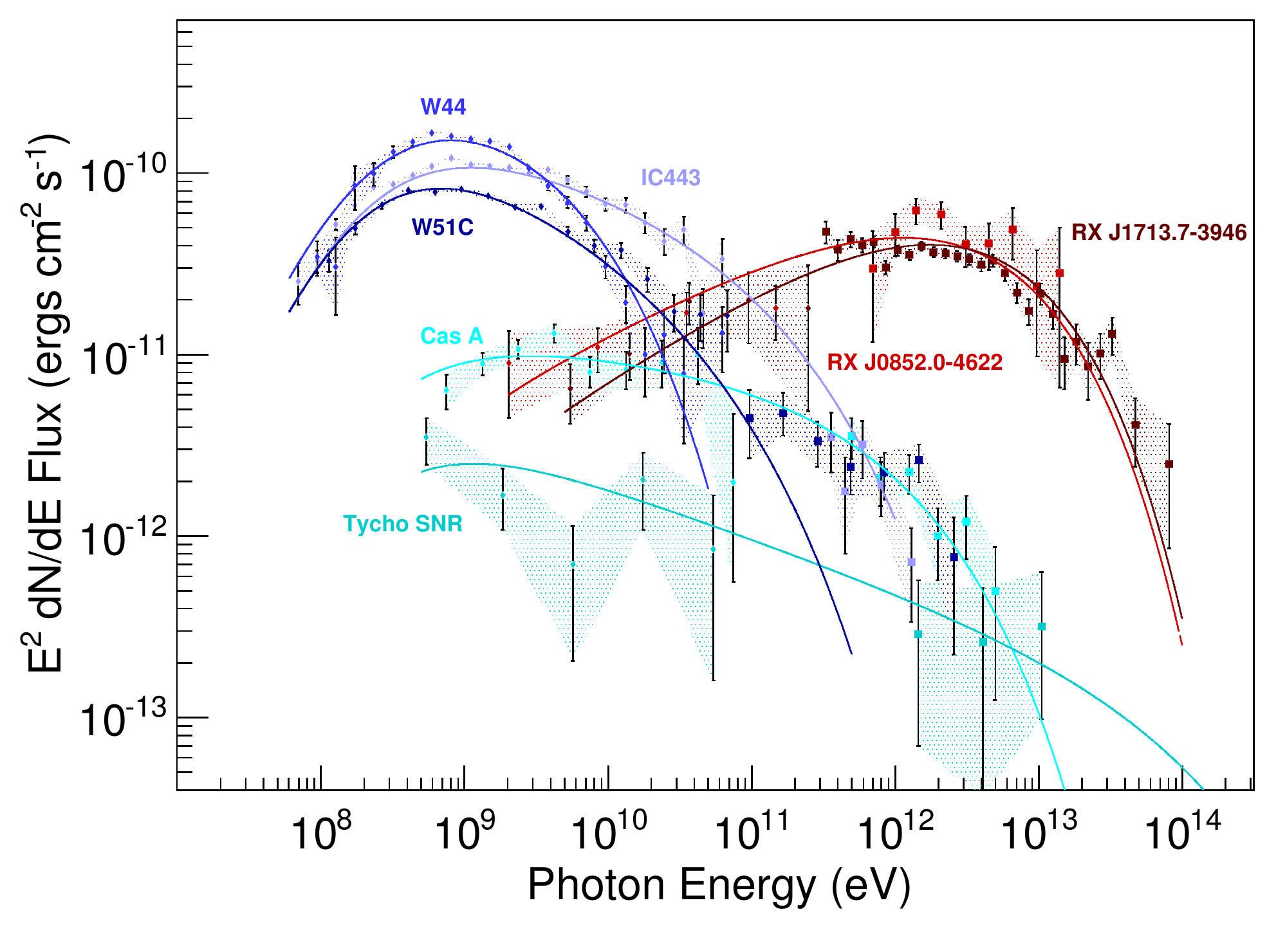}
\caption{Typical gamma-ray energy spectra for several of the most
  prominent SNRs. Young SNRs ($<1000$ years) are shown in cyan. These
  typically show smaller gamma-ray fluxes but rather hard spectra in
  the GeV and TeV band. The older (but still so-called young)
  shell-type SNRs RX\,J1713.7--3946 and RX\,J0852.0--4622 (Vela
  Junior) of ages $\sim 2000$ years are shown in red colors. These
  show very hard spectra in the GeV band ($\Gamma = 1.5$ and a peak in
  the TeV band with an exponential cutoff beyond 10 TeV. The mid-aged
  SNRs ($\sim 20,000$ years) interacting with molecular clouds (W44,
  W51C and IC443) are shown in blue. Also shown are hadronic fits to
  the data (solid lines).}\label{fig::SNRSpectra}
\end{figure}

\begin{figure}[!htb]
\includegraphics[width=\textwidth]{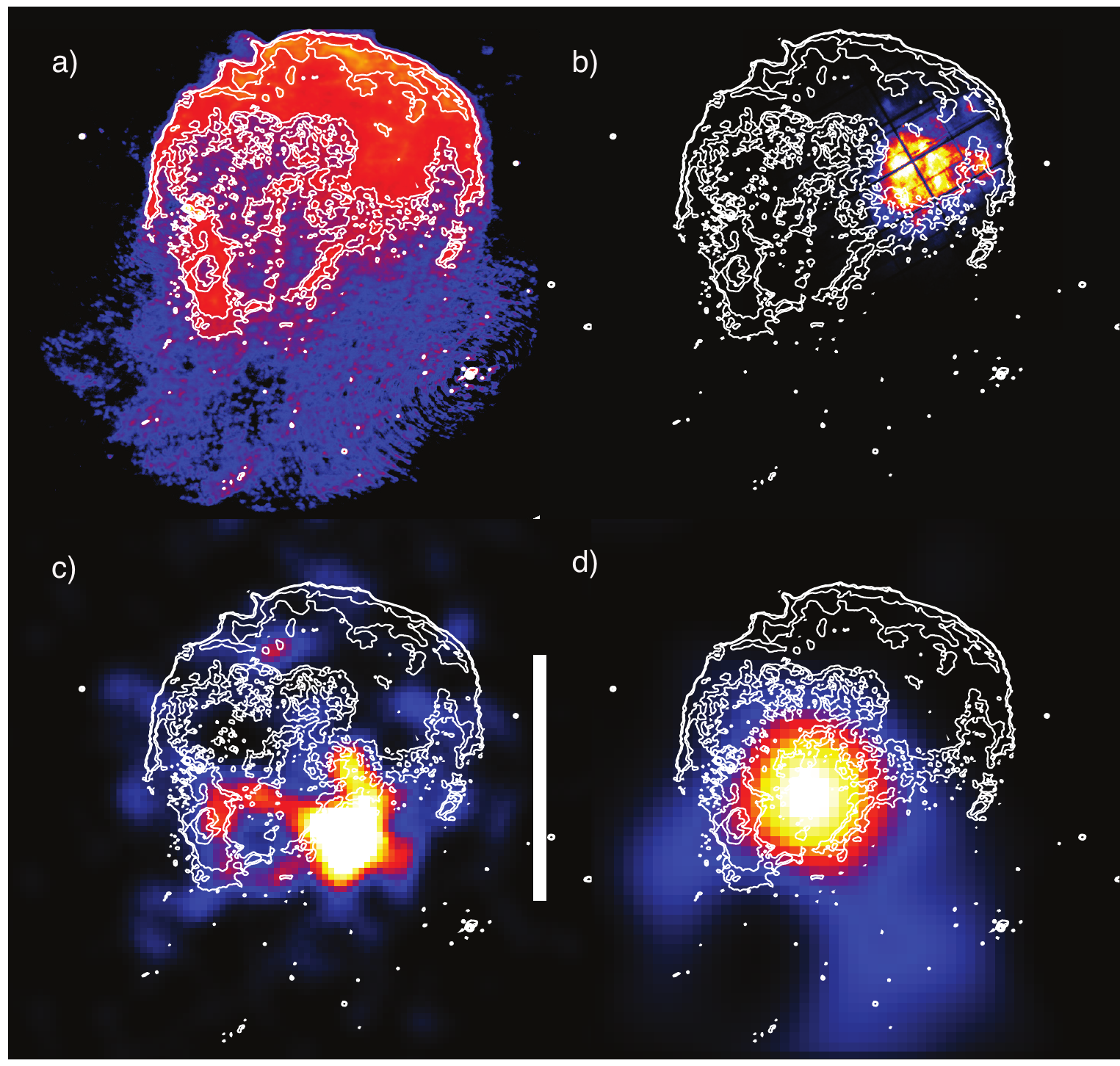}
\caption{Multiwavelength data for SNR IC443 in Galactic coordinates.
  The white bar is 0.4$^\circ$ long. Data are taken from: a) VLA 330
  MHz continuum emission (fits image kindly provided by Dr. Gabriela
  Castelletti)~\cite{IC443Radio}, b) XMM observations (0.3-8 keV)
  \citep{IC443Troia}, c) GeV gamma rays observed by {\emph{Fermi}}-LAT above 10
  GeV (source class, 6.5 years of data) smoothed counts d) MAGIC
  observations above 380 GeV~\cite{IC443MAGIC}. }\label{fig::IC443Map}
\end{figure}

\begin{figure}[!htb]
\includegraphics[width=0.5\textwidth]{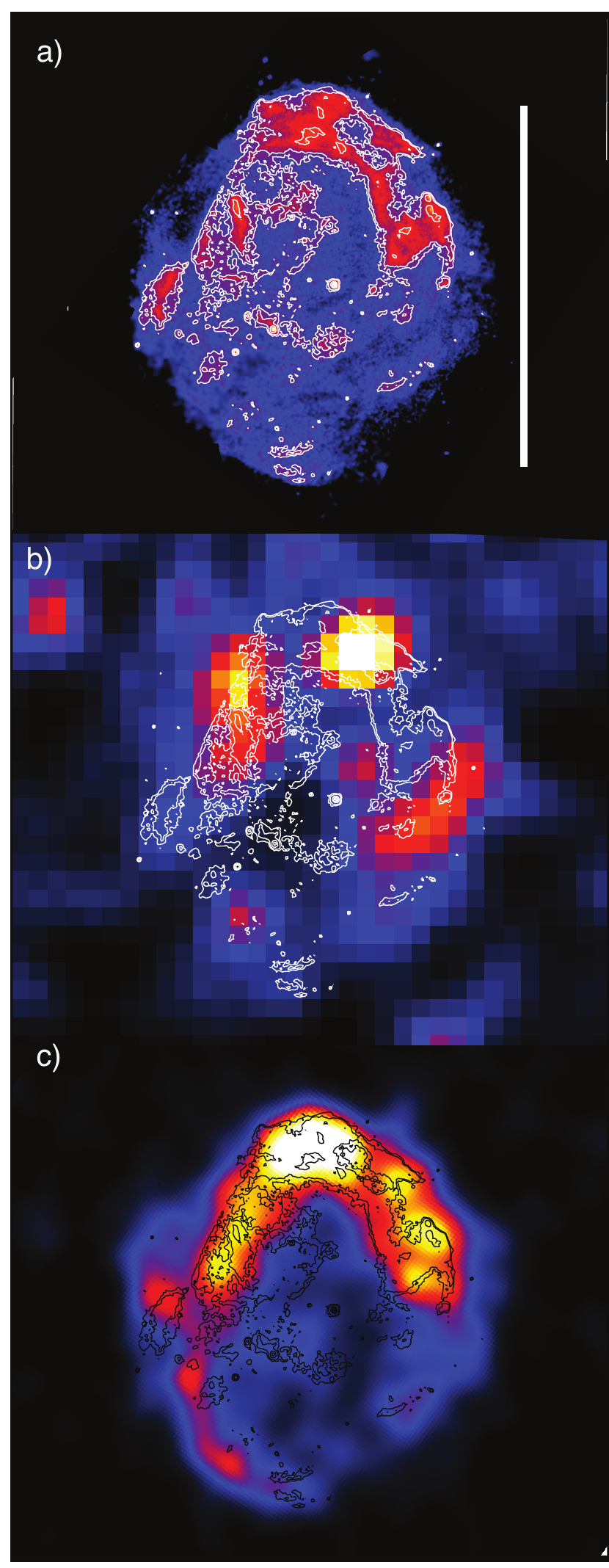}
\caption{Multiwavelength data for SNR RX\,J1713.7-3946 in Galactic
  coordinates.  The white bar is 1$^\circ$ long. Data are taken from:
  a) XMM-Newton observations \cite{AceroXMM}, c) GeV gamma rays
  observed by {\emph{Fermi}}-LAT (source class, 6.5 years) d) H.E.S.S.
  observations~\cite{HESS:rxj1713p3}. }\label{fig::RXJ1713Map}
\end{figure}

Indeed, beyond pulsars and PWNe (which are generally assumed to be
dominated by CR electrons) the largest number of detected gamma-ray
sources in the Galaxy are SNRs.  The {\emph{Fermi}}-LAT team is about
to release its catalog of SNRs in which the data are analysed for each
of the known SNRs~\citep{Greens} in our Galaxy, resulting in
approximately 40 detections. For these detections two classes emerge
(see e.g.~Figure \ref{fig::SNRSpectra}): the largest class in the
GeV-detected SNRs are those known to be interacting with molecular
clouds such as IC443, W44 and W51C (see
Figure~\ref{fig::IC443Map}). The second class are young SNRs that are
typically less luminous at GeV energies, have harder spectra and are
often also detected at TeV energies. At TeV energies 11 shell-type
SNRs are detected, including such objects like Tycho's SNR, Cas A, SN
1006 and RX\,J0852.0--4622 (Vela Junior) as well as RX\,J1713.7--3946
(see Figure~\ref{fig::RXJ1713Map}). The results seem to indicate that
the CR efficiencies $\epsilon_{CR}$ (the efficiency of converting the
SN explosion energy into CRs) are broadly consistent with a value of
10\%, albeit with rather large errors for individual SNRs due to
uncertainties about distance, explosion energy and target density
surrounding the remnants~\cite{Fermi:SNRcat2}. A study at TeV energies
with H.E.S.S. based on the Galactic plane
survey~\citep{HESS:scanpaper1, HESS:scanpaper2} came to similar
conclusions~\citep{HESSPopulationStudySNRs}.

\subsubsection{Supernova remnants interacting with interstellar material}
This is the largest class of GeV-detected objects and the brightest
SNRs IC443, W44 and W51C are the brightest objects of this class on
the GeV sky (see Figure~\ref{fig::SNRSpectra}). The brightness stems
from the rather large density of target material stemming from the
interaction of the shock wave with surrounding molecular clouds (up to
$n = 1000 \mathrm{cm}^3$). For these objects a correlation between the
GeV and the radio flux seems to emerge~\citep{Fermi:SNRCat1}
indicating non-thermal emission from relativistic particles. For IC
443 and W44, the characteristic low-energy cutoff in the energy
spectrum (``pion bump'') has been detected~\citep[][and see
Figure~\ref{fig::SNRSpectra}]{Fermi:PionBump}, clearly demonstrating
that the gamma-ray emission in the GeV band is dominated by
$\pi^0$-decay and therefore unambiguously proving that CRs protons are
accelerated in the shock waves of Supernova remnants. This measurement
provides an important step forward in relating SNRs to the origin of
CRs. The gamma-ray spectra of these (mid-aged, i.e.  $>10,000$years)
objects also show a high-energy break at around 2 GeV (for W44) and 20
GeV (for IC443) well short of the knee in the spectrum of the CRs. At
this point it is unclear, whether this high-energy break is related to
the age of the SNRs, to Alfven damping in a dense
environment~\citep{MalkovDiamond2011}, to the velocity of scattering
centers responsible for the particles diffusion around the
shock~\citep{Blasi2013} (this would change the compression factor) or
to the effects of escape from the remnant~\citep{Gabici2009}.  While
W44 is so far not detected at TeV energies, IC~443 has been detected
with rather soft energy spectrum both by MAGIC~\citep{IC443MAGIC} and
by VERITAS~\citep{IC443VERITAS}. For IC 443 the centroids of the GeV
and TeV emission are not coinciding, a fact that could be explained as
the result of the escape of high energy CRs from the SNR shell and of
their interaction with the ambient medium. In this scenario, the
diffusion coefficient for the multi-TeV CRs is of the order of
$D \approx 10^{27} (\epsilon_{CR}/0.1)^2$ $cm^2/s$ (with
$\epsilon_{CR}$ the efficiency of converting the SN explosion energy
into CRs), a diffusion coefficient that is much smaller than the
average diffusion coefficient in the Galaxy~\citep{Torres2008}.  While
these measurement provide an important proof of the acceleration of CR
protons in shock wave of SNRs, the lessons that can be learned on the
acceleration of the bulk of CRs in our Galaxy are somewhat limited
since the maximum particle energy is way short of the ``knee'' in the
spectrum of CRs. In this respect the second class of objects -- the
young SNRs are much more important.

\subsubsection{Young Supernova remnants}
Young SNRs are the prime candidates for acceleration of protons to
very high energies. If one assumes that shock acceleration is
universal, the emission from these objects into gamma rays and whether
the emission is predominantly leptonic or hadronic in origin is
strongly influenced by the environment (i.e. the density of
interstellar material and of radiation fields in the shock
region). The most prominent of the young SNRs that have been detected
both at GeV and TeV energies are
RXJ\,J1713.7--3946~\citep{HESS:rxj1713p3, Fermi:rxj1713}, Tycho's
SNR~\citep{VERITAS:tycho, Fermi:tycho}, and Cas
A~\citep{HEGRA:casA, VERITAS:casA, MAGIC:casA, Fermi:casA,
  Fermi:casA2}. The situation for RX\,J1713--3946 (and its companion
TeV shell-type SNR RX\,J0852.0--4622) -- arguably the most-discussed
case of a gamma-ray emitting SNR and the brightest object of its class
at TeV energies -- is rather complicated. The spectral shape measured
over a large energy range with {\emph{Fermi}}-LAT and H.E.S.S. from
$\sim 1$GeV to $\sim 100$ TeV suggests a leptonic origin of the
gamma-ray emission due to the rather hard photon index $\Gamma = 1.5$
in the GeV range~\citep[][and see
Figure~\ref{fig::SNRSpectra}]{Fermi:rxj1713}. In addition, the absence
of thermal X-ray emission which is expected in hadronic scenarios,
even for a case in which electron heating proceeds very slowly
suggests very severe upper limits on the density of gas in the shock
region~\cite{Ellison2010}. This suggests that the region lacks the
target material for the interaction of potentially accelerated protons
to produce neutral pions further supporting the leptonic
scenario. However, ~\citep{Zirakashvili} pointed out that the
expectations for the hardness of the spectrum and for the thermal
X-ray emission can change strongly if assuming a non-uniform (clumpy)
structure of gas density around the SNR (as expected if the progenitor
is a massive star in a molecular cloud). In this case the diffusion of
CRs into the clumps might be such that only the highest energy CRs can
penetrate into the center of the clumps and therefore the proton
spectrum ``seen'' by the gas target might be extremely hard and
therefore produce a very hard gamma-ray spectrum~\citep[see also][for
a comprehensive discussion]{GabiciAharonian}. Whether the requirements
on the clumpiness of the medium and the density in the clumps is met
in RX\,J1713.7--3946 is still a matter of active debate.  The number
of clumps in this scenario needs to be sufficiently large (of the
order of 1000) to make the gamma-ray emission spatially smooth. If the
gamma-ray emission is indeed caused by accelerated hadrons the maximum
energy of the protons is $\sim 150$ TeV~\citep{Zirakashvili}. Such
hadronic scenarios can not be ruled out with current data but require
multi-zone modelling of the emission region ~\citep[which in fact fit
the observed parameters rather well as e.g. shown in
][]{GabiciAharonian}. It should be noted that also in the case of
one-zone leptonic models problems exist. The required density of the
IR light necessary to explain the H.E.S.S. data is significantly
(about a factor of $\sim 25$) larger than expected. Also, the one-zone
leptonic model requires magnetic field strengths of the order of
$\sim 10\mu G$, far too low to explain the sharpness of the X-ray
filaments (with electron cooling) and the observed short-term (yearly)
variability of the hard X-ray emission.  So it seems, that also for
the leptonic emission a multi-zone model is necessary to explain all
the observational data. The future will provide more sensitive
observations of the thermal X-ray emission with Astro-H of this
object~\citep{ASTROH:shocks} and ultimately, the detection of
ultra-relativistic neutrinos from this object might be necessary to
unambiguously distinguish between the different scenarios. Clearly,
gamma-ray observations with the Cherenkov Telescope Array (CTA) will
provide significantly higher angular and spectral resolution studies
for this object. From the current observational data it seems clear,
however, that this object is located in a very complex environment and
disentangling the environmental parameters from the acceleration
properties will be a challenging task also in the future.

The cases of Tycho and Cas~A might be clearer (see
Figure~\ref{fig::SNRSpectra}). For these, the gamma-ray spectrum
continues with a spectral index of $\Gamma=2.0$ from the high-energy
TeV range observed by VERITAS~\citep{VERITAS:tycho} into the
lower-energy GeV region, thereby matching more the expectation from a
hadronic scenario and diffusive shock
acceleration~\citep{MorlinoTycho, Blasi2014}. In particular Tycho's
SNR, resulting from a historical Type Ia progenitor supernova
explosion expanding into an undisturbed and therefore more uniform
interstellar medium, seems to be a very good case for the hadronic
interpretation of the broad-band gamma-ray emission. Like in the case
of other SNRs the thin X-ray rims suggest large magnetic fields in the
shock region ($\sim 300 \mu G$) suggesting a maximum energy of
accelerated protons of up to $\sim 500$ TeV~\citep{MorlinoTycho}. A
similar case for hadronic emission can be made for Cas~A, for which
the gamma-ray spectra seem to suggest two energy breaks, one at low
energies ($\sim 1.5$ GeV resulting in a very steep index below that
energy, again consistent with a pion-bump~\citep{Fermi:casA,
  Fermi:casA2}. In the energy range between the {\emph{Fermi}}-LAT
measurements and the measurements of MAGIC~\citep{MAGIC:casA} and
VERITAS~\citep{VERITAS:casA} an additional steepening is necessary. If
this steepening indicates a cutoff in the energy spectrum of the gamma
rays efficient particle acceleration in Cas~A to very large energies
is at least questionable~\citep{Blasi2014}. Recent {\emph{NuSTAR}}
hard X-ray data (3--79 keV) suggest that the GeV emission is spatially
not associated with the bright features in the hard X-ray band while
the TeV emission may be. While the angular resolution of the gamma-ray
data might not be sufficient at this time to unambiguously make this
claim, such a measurement might suggest that in the gamma-ray emission
in Cas~A both hadronic and leptonic emission mechanisms may be at
work~\citep{Nustar:casA}.

An interesting case is the other two young SNRs RCW~86 and SN~1006
(both of age $\sim 1000$ years). Both of them are detected at TeV
energies in deep H.E.S.S. observations~\citep{HESS:rcw86, HESS:sn1006}
and both of them show no emission in the GeV range (in the case of SN
1006) or extremely faint emission (in the case of RCW 86). These
observations put rather stringent constraints on the spectral index of
the protons in a hadronic scenario and limits them to be significantly
harder than $\Gamma = 2$~\citep{Fermi:rcw86, Fermi:sn1006}. In the
case of RCW~86 a recent detection with {\emph{Fermi}}-LAT determines
the spectral index to be $\Gamma = 1.4$~\citep{Fermi:rcw862}. Similar
to the case of RX\,J1713.7--3946, these observations favor a leptonic
model if one assumes a one-zone model with a uniform density. All the
complications described above for RX\,J1713.7--3946 might also apply
to these remnants.

Summarising the situation for young SNRs: while the situation in
RX\,J1713--3946 remains complex probably due to the complex
environment in which the remnant propagates the cases for Tycho SNR
and Cas~A seem to be more clearly indicating predominantly hadronic
emission with typical efficiencies for converting explosion energy
into the acceleration of CRs of the order of 10\%. Measuring
the spectra of these remnants to the very highest energies and
determining the shape and presence of energy cutoffs with CTA will
provide important additional constraints on the acceleration
processes. 
 
\subsubsection{Cosmic rays diffusing away from the sources}

If a molecular cloud is located in the vicinity of the acceleration
region the diffusion of CRs can be studied. Depending on the
spatial location of the molecular cloud and the accelerator different
signatures are expected. The energy dependence of the diffusion
coefficient of the CRs typically alters their spectrum from the source
to the interaction region. This in turn affects the spectral shape of
the gamma-ray emission in the cloud and in the accelerator. Measuring
wide-band spectra of accelerator and cloud can therefore help to
constrain the diffusion coefficient in these regions.

One of the most prominent targets for the study of diffusion of CRs in
our Galaxy is the Cygnus region. Cygnus was detected both at GeV
energies~\citep{FermiCygnus} and at TeV energies~\citep{HEGRACygnus,
  VERITASCygnus, MILAGROCygnus2, ARGOCygnus}. In the
Cygnus region, {\emph{Fermi}}-LAT detected an extended excess of
gamma-ray emission, the so called {\emph{Fermi cocoon}}, which seems
to be related to the combination of many powerful SNR and stellar-wind
shocks. Recently the ARGO Collaboration claimed the association of the
source ARGO J2031+4157 with the cocoon~\citep{ARGOCygnus}, and which
provided a connection between the spectrum of the cocoon measured by
{\emph{Fermi}}-LAT with that of the Milagro source MGRO J2031+41. The
spectral index of the gamma-ray emission when combining these
measurements is $\Gamma = 2.16$ with a cutoff at $E_c = 150$TeV. The
detection of hard extended emission above 1 GeV towards the central
part of Cygnus X is spatially coincident with a cavity in the
interstellar medium blown by the winds and ionisation fronts from
Cyg~OB2, NGC 6910 and other massive stellar
clusters~\citep{FermiCygnus}. The hard spectrum of this emission is
inconsistent with e.g. local CR emissivities and has therefore been
interpreted as the signature of freshly accelerated particles. The
scale of this region is $\sim 50$pc, the total luminosity in gamma
rays above 1 GeV represents however only a modest fraction (0.03\%) of
the mechanical wind power in the stellar winds of Cygnus OB2. This
discovery seems to confirm a long-standing hypothesis that
massive-star forming regions can accelerate particles to relativistic
energies. Additionally, it provides an observational test case to
study the escape of CRs from their sources and the impact of
wind-powered turbulence on their early evolution.

Another important class of observations has been made for regions
surrounding shell-type SNRs, such as W28, and W44. In W44 CRs escaping
from the SNR region and interacting with molecular clouds in the
vicinity of the remnant have been observed in the GeV range with the
{\emph{Fermi}}-LAT~\citep{W44Clouds}. Also for W28 gamma-ray emission
from surrounding clouds has been detected both at
GeV~\citep{W28Fermi2} and at TeV energies~\citep{W28HESS}. The TeV
emission coincides well with the position of the three most massive
molecular clouds. One of these is interacting with the SNR shell
(HESS\,J1801-233) while the other two are located outside of the SNR
(HESS\,J1800-240A and B). A joint fit to all three suggests that the
diffusion coefficient in this region is much smaller (by up to 2
orders of magnitude) than that in the Galaxy as a whole -- a result
consistent with the measurements in IC 443. It should however be
noted, that for all these objects the uncertainty about the age,
distance and three-dimensional relation between the acceleration
region and the clouds makes the conclusions somewhat uncertain. If
indeed the diffusion coefficient is severely suppressed around SNRs,
it might be related to an enhancement in the magnetic turbulence
surrounding the remnant due to CRs streaming away from the source.

Finally, on a larger scale the {\emph{Fermi bubbles}} can be used to
study the diffusion of particles (see discussion below) if indeed they
are related to hot outflows powered by activity of the supermassive
black hole in the Galactic center~\citep{Yang2013, Cheng2011,
  Dogiel2014, Chernyshov2011, Carretti2013, Selig2014, Crocker2014}.

\subsubsection{Maximum energy in Cosmic rays}

The maximum energy attainable in SNR shocks is directly related to the
magnetic fields that confines the particles at high energies to the
shock area In a recent
investigation~\citep{SchureBell2013} assumed that escaping CRs trigger
magnetic field amplification through streaming instabilities. They
estimated the number of particles escaping as a function of shock
velocity and SN type (ambient density). They concluded that young SNRs
reach $\sim 200$TeV energies for shock velocities of
$v_{sh} = 5000 \mathrm{km/s}$. Using the observed properties of Tycho,
SN 1006 and Cas A, the authors determine $E_{\mathrm{max}}$ to be 100
TeV, $<60$ TeV, and 280 TeV respectively, largely consistent with the
gamma-ray observations.  The authors conclude that the highest
energies for CRs are reached in the very early phases of SNR
evolution, especially in the core-collapse SNe in a dense wind, which
are representative for the early stages of most type II SNe. PeV
energies are then reached only in the first few hundred years and only
in a dense wind environment.  For the known Galactic SNRs, Cas~A seems
to be the best candidate to have been an accelerator to the knee,
although only in the early stages of its evolution and now no more.
It seems plausible that younger SNRs with even higher shock velocities
can accelerate particles up to the knee -- however, the question is
whether these SNRs have swept up enough material and accelerated
enough material to account for the total flux of CRs seen at
Earth. Nevertheless, the absence of a Pevatron at this time does not
seem to pose a major problem to the paradigm of diffusive shock
acceleration in SNR shocks, given that these Pevatrons possibly
operate for only a very short period of
time~\citep{GabiciAharonian2007, BlasiAmatoCaprioli2007}.

Interestingly, there seems to be a second population of CRs in SNRs,
those that do not have enough energies to escape before the end of the
lifetime of the shock wave. CR with energies up to $\sim 200$ GeV are
confined to the SNR for the lifetime of the SNR and form a CR bubble
that gets released into the Galaxy only at the end of the SNR's
life~\citep[see e.g.][]{Bell2014}. These bubbles might not be strong
emitters of gamma rays due to the low gas densities in the bubbles.
The recently discovered {\emph{Fermi}} bubbles might be related to this
phenomenon.

\subsubsection{Cosmic rays in other galaxies}

We know that the GeV gamma-ray emission of our Galaxy is dominated by
the interaction of CRs with interstellar material. Whether this is
true also for other galaxies remains a matter of active study. So far,
also the Large Magellanic Cloud (LMC)~\citep{Fermi:lmc} and the Small
Magellanic Cloud (SMC)~\citep{Fermi:smc}, M31~\citep{Fermi:m31} as
well as several active star forming Galaxies (most notably M~82 and
NGC~253)~\citep{Fermi:ngc253, Fermi:ngc2532} have been detected at GeV
and TeV energies. For NGC 253 the gamma-ray spectrum can be described
by a power law with photon index $\Gamma = 2.2$ from 200~MeV to 5
TeV~\citep{Fermi:ngc253, HESS:ngc253, HESS:ngc2532}.  Active star
formation leads to an enhanced rate of Supernova explosions and
therefore to an enhanced CR density. Basic considerations suggest that
the timescales for the acceleration of CRs and subsequent energy loss
(via pion-decay, or advection by bulk outflows) are shorter than the
timescales of starburst activity in Galaxies. Consequently, a balance
is expected between energy gains and losses for Galactic CRs in a
burst of star formation. In a population study of 64 close-by galaxies
selected by their abundant molecular gas, a 3-year data analysis of
{\emph{Fermi}}-LAT data revealed a clear (quasi-linear) scaling
relation between gamma-ray luminosity and both radio continuum
luminosity and total infrared luminosity. In the paradigm that SNRs
channel approximately 10\% of their mechanical energy into CR nuclei
with kinetic energies $> 1$ GeV, the normalisation of the observed
scaling relation between gamma-ray luminosity and star-forming rate
implies that starburst galaxies have an average calorimetric
efficiency for CR nuclei of 30–-50\% if the gamma-ray emission is
dominated by neutral pion decay~\citep{Fermi:ngc2532}. Interestingly,
the rather hard gamma-ray spectra ($\Gamma = 2.2–-2.3$) of starburst
galaxies such as NGC 253~\citep{HESS:ngc2532} and
M82~\citep{VERITAS:m82} suggest energy-independent loss mechanisms for
CR nuclei in starbursts, as opposed to the diffusive losses which
likely shape the observed gamma-ray spectra of the Milky way and the
other quiescent Local Group galaxies (such as the LMC, SMC). These
energy-independent loss mechanisms could include hadronic interactions
but also advective transport of CR nuclei in starburst galaxies. The
overall normalisation of the relation between gamma-ray luminosity and
star formation rate ultimately supports the understanding that the
majority of CR energy in most galaxies eventually escapes into
intergalactic space losing only a few percent of their energy through
pion production and ionisation in Milky Way-like
galaxies~\citep{Strong2010}. Star-forming galaxies have more gas and
possibly heavier losses, yet for the detected objects galaxies M82 and
NGC 253 still 60-80\% escape (likely in a wind)~\citep[see
e.g.][]{Lacki2011, HESS:ngc2532, Fermi:ngc2532} into the intergalactic
medium. Here they build up over the Gyrs of cosmic star
formation. ~\citep{Lacki2015} has shown, that CR can couple to the
intergalactic medium and its weak magnetic field and ultimately can
reach a density within a factor of a few of the Ly$\alpha$ forest
pressure, probably having the strongest effect at redshifts of
$z \sim 1$ and larger, when the star-formation was largest and most of
the intergalactic medium was cool.

\subsection{The Search for Dark Matter}
Because of the quadratic dependence of the self-annihilation rate on
the dark matter density the detectability of any particular region in
the Universe (see equation~\ref{eq::annihilation}) strongly depends on
the density distribution of dark matter particles along the line of
sight (so-called J-factor). Unfortunately, dark matter densities are
not very well constrained by numerical simulations, especially in the
innermost regions of Galaxies. In fact, simulations originally showed
that the collapse of cold dark matter gives rise to rather cuspy dark
matter haloes (something that would favor the indirect detection of
dark matter because of the $\rho_\mathrm{DM}^2$ dependency). On the
other hand, observations of galaxy rotation curves favor constant
density cores (so-called `cusp-core problem'',\citep[see
e.g.][]{FloresPrimack1994, NFW1996, Moore1999}. An
additional complication stems from substructure in the dark matter
distribution that is currently not resolved in cold dark matter N-body
simulations, (i.e. below $\sim10^5 M_{\odot}$). This unresolved
substructure can have a very large impact, in particular in objects
such as galaxy clusters. Since substructure will further enhance the
annihilation signal this effect is typically quantified in terms of
the so-called boost factor $B$ defined as the ratio of the true
line-of-sight integral to the one obtained when assuming a smooth
component without substructure. Finally, the situation is further
complicated by the fact that for many objects (such as e.g.\ the Milky
Way) baryonic matter dominates the inner parts of the gravitational
potential. Baryons are expected to have a significant impact on the
dark matter profile compared to the numerical simulations which are
generally dark matter-only. The infall of baryons is expected to alter
the inner dark matter profiles. The profile could either be steepened
through adiabatic contraction~\citep{Jesseit2002,
  Prada2004}, or it could be flattened through the occurrence of
repeated star bursts triggered by baryonic infall which tends to
render the gravitational potential shallower since the star burst
activity drives out the baryons from the inner
parts~\citep{Mashchenko2008, Governato2012}.

The choice of the assumed profile of the density distribution
constitutes therefore one of the prime uncertainties in studying dark
matter using gamma rays. The resulting uncertainties on the expected
flux limits for individual dwarf spheroidals are between a factor of 3
for well-constrained objects like Sculptor up to a factor of 10 for
objects such as Coma Berenices. For the center of our Galaxy, arguably
the most promising target in terms of expected gamma-ray flux from
dark matter annihilations, these uncertainties are considerably
larger. \citep[See][]{CatenaUlio2010} for a discussion on the dark
matter profile in the inner Galaxy from a meta-analysis of kinematic
data of the Milky way. Estimates can differ by up to a factor of 50
depending on the choice of the profile. For clusters of galaxies the
main uncertainties come from the treatment of substructure below the
resolution limit of current numerical simulations. Also here, the
uncertainties in the J-factor can be several orders of magnitude, and
the dark matter profile itself can be severely modified in these
objects by the presence of substructures.

No unambiguous evidence of gamma rays from dark matter annihilation
has been detected so far (see a discussion of the Galactic center
later). Therefore, upper limits on the annihilation cross sections
(under the assumptions of certain dark matter profiles) have been
derived for various objects. Interestingly, many of these upper limits
especially in the GeV range already touch on physically meaningful
parts of the dark matter parameter space. In particular reaching the
thermal relic cross section in the stacking of dwarf spheroidal
Galaxies with the {\emph{Fermi}}-LAT should be noted. The various sources for
which upper limits have been derived will be discussed in the
following:

\subsubsection{Dwarf Spheroidals}

In contrast to the aforementioned uncertainties in the inner parts of
the density profile of objects on Galaxy or galaxy cluster scales,
dwarf spheroidal galaxies can represent a very clean system to search
for dark matter annihilation. Indeed, star formation is usually very
much suppressed in these objects, so astrophysical foregrounds that
produce gamma rays are less of an issue in these objects. 
Boost factors through substructure below the resolution of numerical
simulations are expected to be irrelevant in these objects and
therefore do not add a large uncertainty. Uncertainties related to the
shape of the dark matter profile are generally integrated over and are
at the 10-50\% level. Currently, there are roughly 25 known dwarf
satellite galaxies to the Milky Way and ground-based instruments such
as H.E.S.S., MAGIC and VERITAS as well as the {\emph{Fermi}}-LAT are
actively observing these objects. While none of the objects are
detected with the current generation of gamma-ray instruments,
important conclusions on the properties of dark matter particles can
be drawn from these objects. In particular, a combined analysis of all
known dwarf satellites with the {\emph{Fermi}}-LAT have pushed, for
the first time the annihilation interaction rate limits below the
canonical thermal relic production cross-section of
$3\times 10^{-26} \mathrm{cm}^3 \mathrm{s}^{-1}$ for a range of WIMP
masses (around 10~GeV) for the annihilation into
$\mathrm{b}\bar{\mathrm{b}}$, which often acts as a
benchmark~\citep{FermiDwarfs, GeringerSamethDwarfs}. This
statement holds also if uncertainties in the J-factors for these
objects are included. Given the all-sky capability of the
{\emph{Fermi}}-LAT, a combined analysis of these objects will remain
the cleanest target in the future where more dwarfs are expected to be
detected with future optical surveys such as
Pan-STARRS~\citep{PanStarrs}, DES~\citep{DES} and
LSST~\citep{LSST}. Estimates show that DES might discover 19 to 37 new
dwarf galaxies during the duration of the {\emph{Fermi}}-LAT
mission~\citep{Tollerud2008}. At higher energies ground-based IACTs
have also observed dwarf spheroidals but have not found a
signal~\citep{VERITASDwarfs, HESSDwarfs, MAGICDwarfs}. Their limits
for high WIMP masses are typically several orders of magnitudes away
from the thermal relic interaction rate and are therefore not (yet)
competitive with limits from the {\emph{Fermi}}-LAT at lower energies.

\subsubsection{Galaxy Clusters}

Galaxy clusters are more distant than dwarf spheroidal galaxies or any
of the other targets that are generally used for dark matter studies
using gamma rays. However, similar to the case of dwarf spheroidals,
they are expected to be dark matter dominated. The range of proposed
boost factors due to unresolved dark matter substructure can be
large. Depending on the assumption about the substructure galaxy
clusters become competitive in their expected annihilation signal with
dwarf spheroidals only at the extreme (high) end of boost factors. The
best candidate are massive nearby clusters such as Virgo, Fornax or
Coma~\citep{Pinzke11, SanchezConde, GAO11}. One complication for a
possible detection is that galaxy clusters are also expected to
contain a significant number of astrophysical sources of gamma rays,
such as Active Galactic Nuclei (AGN) or radio galaxies. In addition,
these objects are expected to harbor a significant population of CRs
which should radiate gamma rays through interaction with hadronic
material and subsequent pion-decay. Fermi-LAT has not detected any
signal from a cluster of galaxies, and as long as no signal is found,
ignoring the CR contribution represents a conservative assumption and
is therefore justified~\citep{FermiGalaxyClustersII, Huang, Nezri,
  SanchezConde}.
At higher energies ground-based instruments have pushed for rather
stringent gamma-ray flux limits on galaxy clusters (e.g.\ the MAGIC
telescopes for the Perseus cluster~\citep{MAGICPerseus}, the VERITAS
array for the Coma cluster~\citep{VERITASComa}, and the H.E.S.S.\ array
for the Fornax cluster~\citep{HESSFornax}). However, when making
conservative assumptions about boost factors in these objects, the
limits on the benchmark $\mathrm{b}\bar{\mathrm{b}}$ annihilation
channel are several orders of magnitude away from the canonical
thermal relic interaction rate.

\subsubsection{Isotropic diffuse emission}
The {\emph{Fermi}}-LAT has provided a measurement of a faint diffuse
isotropic signal (IGRB) from 100~MeV to
820~GeV~\citep{Fermi:egb2}. Due to the large-scale structure of the
signal and the inability to reject the electron background, this
emission is not detectable with ground-based detectors. The isotropic
emission exhibits a power law with index $\Gamma = 2.32 \pm 0.02$ with
a significant exponential cutoff at $279 \pm 52$ GeV -- a statement
that is robust against changes in the Galactic diffuse foreground
model. At an energy of 100 GeV, roughly half of the total IGRB
intensity has now been resolved into individual sources by the LAT,
predominantly blazars of the BL Lacertae type. The cutoff in the
energy spectrum can in principle be explained by a single dominant
extragalactic source population with EBL
attenuation~\citep{Fermi:egb2, InoueIoka2012,
  Murase2012}. This signal is expected to contain a contribution of
mainly extragalactic unresolved (sub-threshold) sources combined with
potentially truly diffuse emission.  It is thus possible that the IGRB
emission contains the signature of some of the most powerful and
interesting phenomena in astroparticle physics. Intergalactic shocks
produced by the assembly of Large Scale
Structures~\citep{LoebWaxman2000, Miniati2002, GabiciBlasi2003},
gamma-ray emission from galaxy clusters~\citep{BerringtonDermer2003,
  Pfrommer2008}, emission from starburst and normal
galaxies~\citep{PavlidouFields2002, Thompson2007}, are among the most
likely candidates for the generation of diffuse GeV emission. In
addition, a signal from dark matter annihilation could be imprinted in
the IGRB. While it would be extremely difficult to detect a dark
matter contribution in the IGRB, upper bounds on dark matter
annihilation can be readily derived. The most conservative approach
when calculating upper limits on the dark matter annihilation
interaction rate is to assume that all of the IGRB is caused by dark
matter annihilation. When making rather conservative assumptions about
the contribution of source populations to the IGRB dark matter
annihilation interaction rate limits can be derived~\citep{FermiIGRB,
  KevorkIGRB, BringmannIGRB, HooperIGRB} that are competitive with
other methods, such as dwarf spheroidal galaxies. Obviously, these
limits can be significantly tightened when including additional source
populations - however, the degree to which the contribution from such
classes can be determined is questionable.

The statistical properties of the IGRB additionally encodes
information about the origin of this emission. Unresolved sources are
expected to induce a different level of small-scale anisotropies
compared to truly diffuse contributions. A study of the angular power
spectrum of the diffuse emission at Galactic latitudes $|b| >
30^{\circ}$ between 1 and 50 GeV revealed angular power above the
photon noise level at multipoles $l > 155$ independent of
energy~\citep{FermiIGRBAnisotropy}. The scale independence of the
signal suggests that the IGRB originates from one or more unclustered
populations of point sources.  The absence of a strong energy
dependence suggests that a single source class that provides a
constant fractional contribution to the intensity of the IGRB over the
energy range considered may provide the dominant contribution to the
anisotropy. Recently it has been suggested that a strong
correlation between cosmic shear (as measured by galaxy surveys like
DES and Euclid) and the anisotropies in the IGRB might add an
additional handle on the contribution of dark matter annihilation to
the IGRB~\citep{IGRBCosmicShear}.

\begin{figure}[!htb]
\centering
\includegraphics[width=0.8\textwidth]{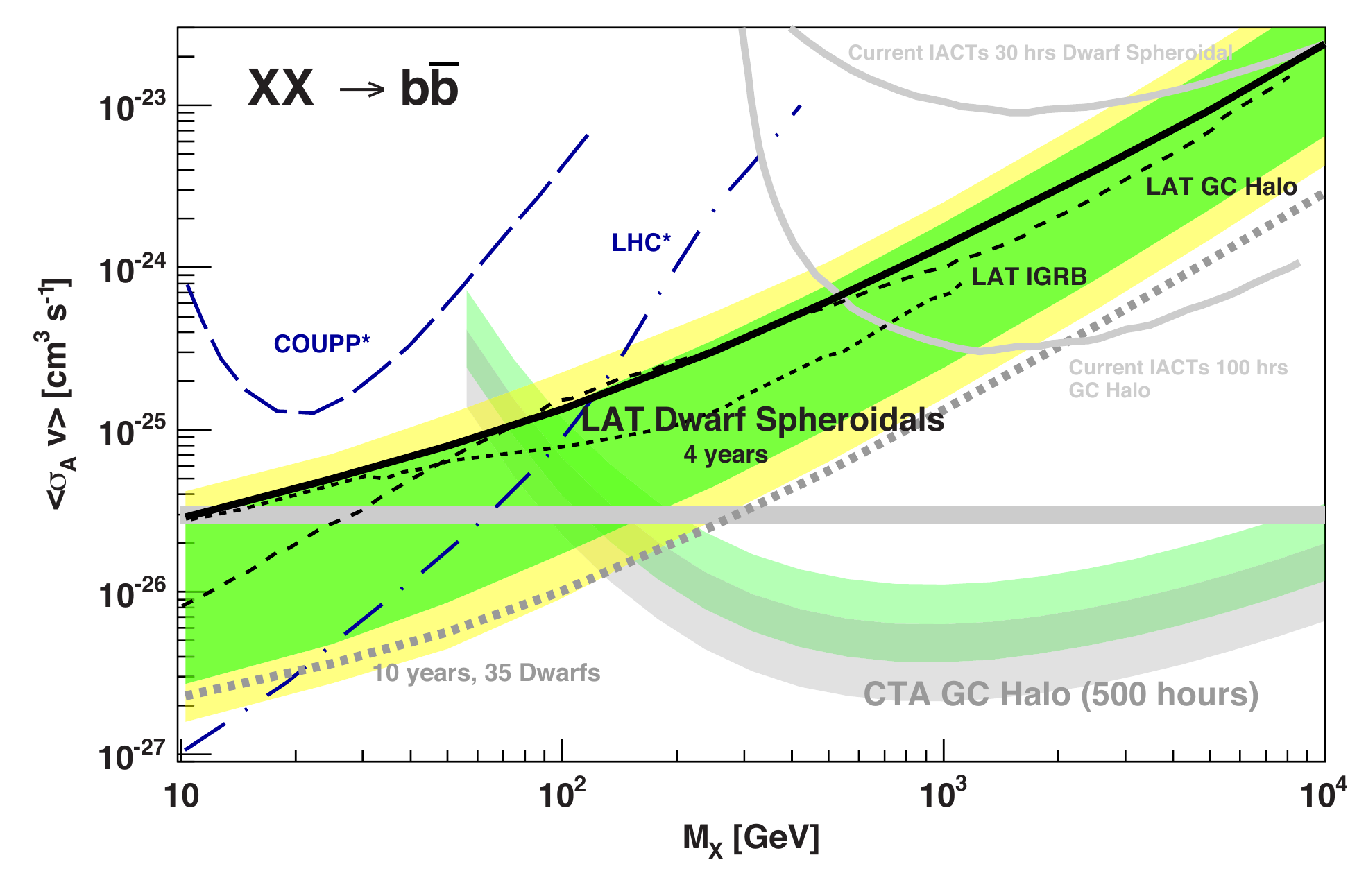}
\caption{Reproduced from~\citep{DMReviewFunk}.Compilation of current
  limits from the {\emph{Fermi}}-LAT on dark matter annihilation interaction
  rate $\sigma v$ using gamma rays, along with projected future limits
  from the Cherenkov Telescope array (CTA). Also shown are limits from
  direct detection (COUPP~\citep{COUPP}) and accelerator experiments
  (LHC ATLAS~\citep{ATLAS} and CMS~\citep{CMS}) transformed to limits
  on the annihilation cross section on the assumption of four-particle
  contact interactions as in ~\citep{Arrenberg}.  The {\emph{Fermi}}-LAT limits
  were taken from the {\emph{Fermi}}-LAT paper on dwarf
  spheroidals~\citep{FermiDwarfs} (solid black), on the isotropic
  diffuse~\citep{FermiIGRB} (dotted), on the Galactic center halo for
  an NFW profile~\citep{FermiHalo} (dashed).  The yellow (95\%) and
  green (68\%) bands give the range of expected limits when repeating
  the procedure multiple times in Monte-Carlo simulations on dwarf
  spheroidals without the inclusion of a signal.  The light green
  (Einasto) and light gray (NFW) band give the expected CTA
  sensitivity (500 hours) for the Galactic center halo under the
  assumption of two different dark matter profiles.  For these CTA
  estimates, the contribution by the US groups (doubling the number of
  mid-sized telescopes) was taken into account. For the Einasto halo
  model the shape parameter was fixed to 0.17. For both models the
  density was normalized to 0.4 GeV cm$^{-3}$ at the solar radius. For
  the sensitivity curves, the signal was evaluated by integrating the
  product of the gamma-ray acceptance and the differential DM flux
  over an annulus of 0.3 - 1.0 deg from the GC. The background in the
  signal region was calculated from the product of the signal region
  solid angle and an energy-dependent model for the spatial density of
  protons and electrons that survive background rejection cuts.  In
  addition, the uncertainty on the background was modeled by assuming
  the existence of a control region with a solid angle five times that
  of the signal region. }
\label{fig::DMLimits} 
\end{figure}

Figure~\ref{fig::DMLimits} summarises the current status of dark
matter limits from various objects.

\section{Future prospects}

The supplemental material gives a detailed account of one of the most
exciting regions on the sky: the Galactic center and its complex
interplay between cosmic ray sources, gamma-rays from propagating
cosmic rays, and dark matter. As described there, given the complexity
of the emission region it is difficult to draw unambiguous conclusions
and future observations will be needed to improve the picture.  In the
absence of a space-mission that would improve the overall sensitivity
over the {\emph{Fermi}}-LAT by an order of magnitude in the near
future the community is looking toward the next generation
ground-based instrument as the next big step in gamma-ray
astrophysics\footnote{it should be noted, that the planned
  Russian/Italian space-mission {\emph{Gamma-400}}~\citep{Gamma400}
  aims to significantly improve the angular and energy resolution
  which might be relevant for line searches}. The Cherenkov Telescope
Array (CTA) is expected to start operation in 2017 with a partially
completed array (current start of construction planned for 2016). It
will provide sensitive observations of the gamma-ray sky over the
energy range from a few tens of GeV to 100s of TeV. To achieve the
optimal sensitivity over that wide a range in energy, CTA will employ
three different telescope sizes: Large Size Telescope (LST, 23 m
diameter), Medium Size Telescope (MST, 10-12 m) and Small Size
Telescope (SST, 4-6 m). The design goal is a point-source sensitivity
of at least an order of magnitude better than currently operating
instrument at the sweet-spot of 1 TeV and a significantly improved
angular resolution, improving with energy from 0.1$^\circ$ at 100 GeV
to better than 0.03$^\circ$ at energies above 1 TeV~\citep{CTA}.

Even though {\emph{Fermi}}-LAT has now unambiguously demonstrated that
protons are accelerated in the shock waves of SNRs, many questions
about the origin of CRs remain open: the maximum energy imparted on
particles in shock waves, the magnification of the ambient magnetic
field to achieve those high energies, the overall efficiency of
converting explosion energy into the acceleration of particles. CTA
with its enhanced sensitivity and angular resolution will add valuable
information to this puzzle. The HAWC detector is about to be completed
and will survey the sky continuously at the highest gamma-ray energies
(around 10--100 TeV) with particular sensitivity to hard spectrum
extended sources. Additionally, neutrino detectors, such as
IceCube~\citep{IceCube} are now starting to detect astrophysical
neutrinos~\citep{IceCubeDetection} at very high energies above
$\sim 100$ TeV. While the current data set is consistent with a
diffuse isotropic flux of neutrinos, pinpointing individual sources
will significantly improve our understanding of CR acceleration at the
highest energies and should be possible with the next-generation
in-ice or mediteranean detector~\citep{IceCubeGen2}.

Gamma rays are sensitive to almost any annihilation channel with a
sensitivity that is closely related to the total annihilation cross
section of dark matter that underlies its total relic abundance today.
Already now, gamma-ray limits are probing below the thermal relic
interaction rate for some of the preferred WIMP mass range. While
direct detection experiments mostly have to deal with uncertainties in
the background estimations, the indirect detection technique is
largely dominated by astrophysical uncertainties. For dwarf spheroidal
galaxies, these are largely mitigated by the lack of astrophysical
backgrounds and tight constrains on the halo profile from dynamical
measurements. For the Galactic center these uncertainties are the
largest but the prospects for a detection are still the highest. A
positive (and credible) detection would entail either a gamma-ray
line, as in the case of the aforementioned one at 130~GeV in the
{\emph{Fermi}}-LAT data or alternatively, from the measurement of identical
spectra that are compatible with a dark matter origin from more than
one source. None of the proposed dark matter search methods (direct
detection, indirect detection or accelerator searches) will be able to
unambiguously claim the detection of dark matter and thus all the
methods are crucial in a viable dark matter program for the future.
Each potential signal will potentially be created by a new (previously
unknown) background -- even in the case of accelerator searches. One
big advantage of the indirect detection techniques is that if a signal
is found in an accelerator or in a direct detector, gamma-ray
measurements will provide the only way to connect the laboratory to
the actual distribution of dark matter on the sky and identify the
nature of the particle through the details of the annihilation
process. In fact detecting a signal from the Galactic center would
allow to measure the dark matter density profile and feed back to
cosmological simulations. In addition, there is a unique region of the
WIMP parameter space that CTA can best address in the near future --
the high-mass ($\sim 1$ TeV) scenario~\citep{CahillRowley}.

The field still has lots of open question that we would like to answer
in the future. Where are the Pevatrons? What is the CR content in
supernova remnants or in Galaxy clusters? Up to what energies do GRBs
accelerate particles? Can giant pair halos around AGNs be detected?
What is the particle nature of dark matter, is the WIMP scenario valid
and if yes, what is the mass and annihilation cross section of this
particle? Clearly for many of these questions the scientific output
from the community would be enhanced if the mission of the
{\emph{Fermi}}-LAT was further extended to provide overlap with HAWC
and with CTA to cover sources over as broad an energy range as
possible.

\begin{summary}[SUMMARY POINTS]
\begin{enumerate}
\item Gamma-ray observations both from space and from the ground
  provide an unprecedented detail of the high-energy Universe.
\item While the {\emph{Fermi}}-LAT has demonstrated the acceleration of protons
  in SNRs, the hunt for the origin of Galactic CRs (meaning the
  dominant source of CRs and the acceleration up to 10$^{15}$ eV is
  still not finally resolved.
\item Young SNRs provide the best candidates for acceleration up to
  the knee.
\item Indirect searches for dark matter using gamma rays complement
  and enhance accelerator and direct search efforts.
\end{enumerate}
\end{summary}

\section*{DISCLOSURE STATEMENT}
The authors are not aware of any affiliations, memberships, funding,
or financial holdings that might be perceived as affecting the
objectivity of this review.

\section*{ACKNOWLEDGMENTS}
I would like to thank everyone who was involved in the preparation of
this manuscript, including Elliott Bloom, Roger Blandford, Seth Digel,
Manuel Kraus, Luigi Tibaldo, David Thompson,  and Matthew Wood. Also
the support of the anonymous referee is gratefully acknowledged.

\newpage

\section{Supplemental Material}

\subsection{Use case: the Galactic Center region}

The Galactic center is arguably one of the most interesting regions on
the gamma-ray sky. When viewed in radio \citep[see
e.g.][]{LaRosa2000}, the central few 100 pc of the Milky way shows a
complex and very active region, suggesting particle acceleration to
very high energies. It is expected to harbor a large number of
potential gamma-ray emitters both in the GeV and TeV region. A large
population of milisecond pulsars, radio pulsars and SNRs is known to
be located within the inner parsec of our Galaxy~\citep{GC:Eisenhauer05}.
The density of gas and radiation fields for the interaction of CR
protons and electrons in this inner region is large. The very center
of our Galaxy can be expected to accelerate particles to relativistic
energies either through a hypothetical termination shock of a wind
from the supermassive black hole \citep[see e.g.][]{AtoyanDermer2004} or
the combined effect from stellar wind shocks~\citep{QuataertLoeb2005},
thereby additionally contributing to the gamma-ray flux from the
region. On top of all these astrophysical particle accelerators the
Galactic center region is expected to be the brightest source of dark
matter annihilation gamma rays on the sky by at least two orders of
magnitude.

\begin{figure}[!htb]
\centering
\includegraphics[width=0.88\textwidth]{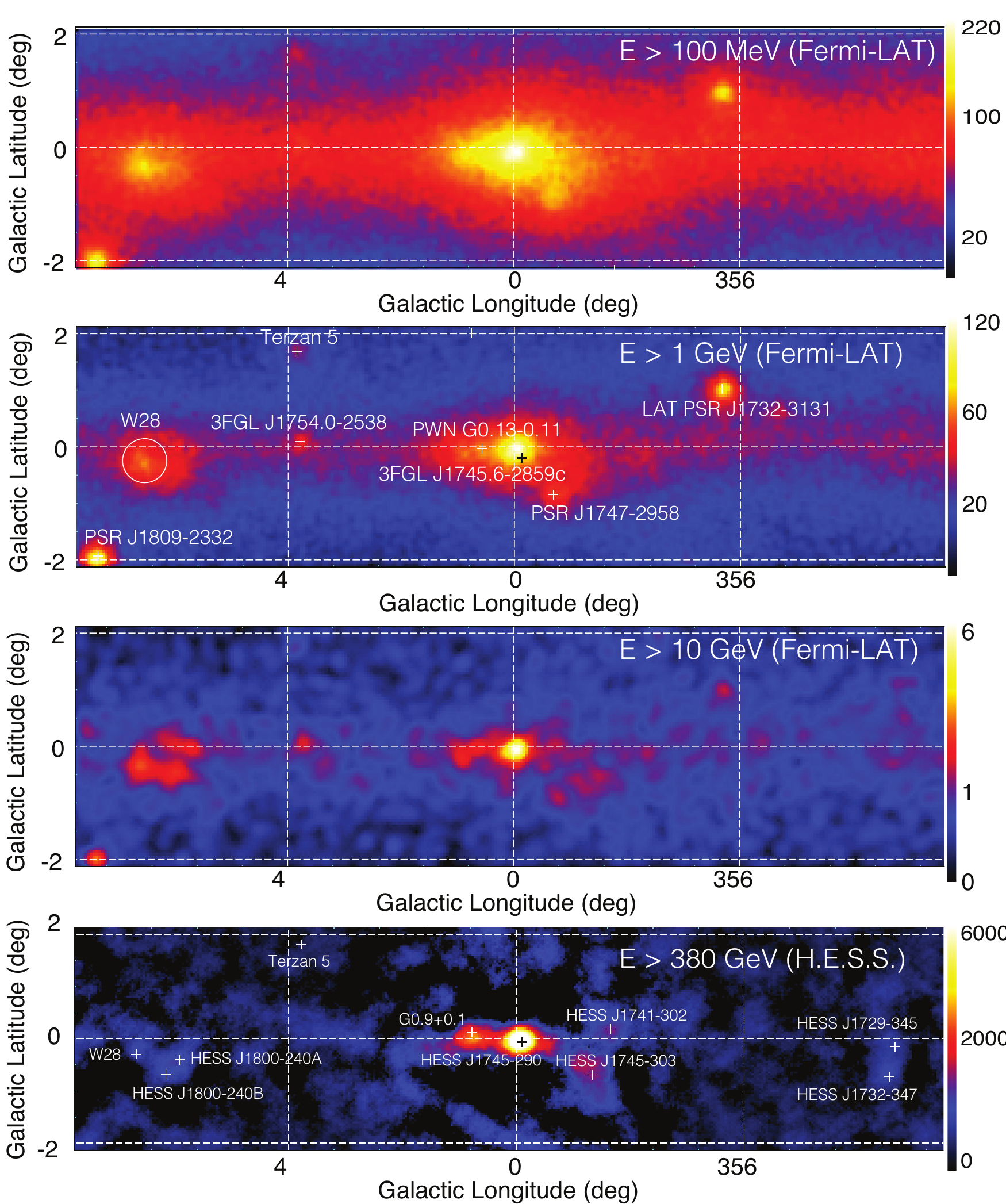}
\caption{Region around the GC ($|l|<8^{\circ}$, $|b| < 2^{\circ}$) in
  gamma rays for different minimum energies. The region corresponds to
  a size of $1100 \times 280$ pc. The top three panels are smoothed
  gamma-ray count maps (top two: Gaussian of width 0.1$^{\circ}$,
  third: Gaussian of width 0.2$^{\circ}$) as observed with the
  {\emph{Fermi}}-LAT for the ultraclean class and 6.5 years of data
  (August 2008 -- February 2015). The pixel size is $0.05^{\circ}$,
  the z-scale shows counts per pixel in a square root scaling. Several
  prominent sources from 3FGL catalog are marked in the second panel
  illustrating the variety of sources. It should be noted that a
  significant number of sources from 3 FGL were not marked in this
  panel. The fourth panel shows a H.E.S.S. excess map from the
  Galactic plane survey as published for the 2004 data
  in~\citep{HESS:scanpaper2} for hard cuts above 380 GeV. The
  TeV-detected sources from TeVCat webpage
  (http://tevcat.uchicago.edu) are marked, although it should be
  stressed that only 1 year of data was used in the map and certain
  areas of this region received a significantly larger amount of
  observations in subsequent years.}
\label{fig::GCMaps} 
\end{figure}

In the following I will focus on the inner 5$^\circ$ from the Galactic
center and describe the observational data and the conclusions that
can be drawn from the observations of that region. An observational
overview of the region is given in Figure~\ref{fig::GCMaps} and
Figure~\ref{fig::GCSpectra}.

\subsubsection{The Central gamma-ray source} 
\begin{figure}[!htb]
\centering
\includegraphics[width=0.8\textwidth]{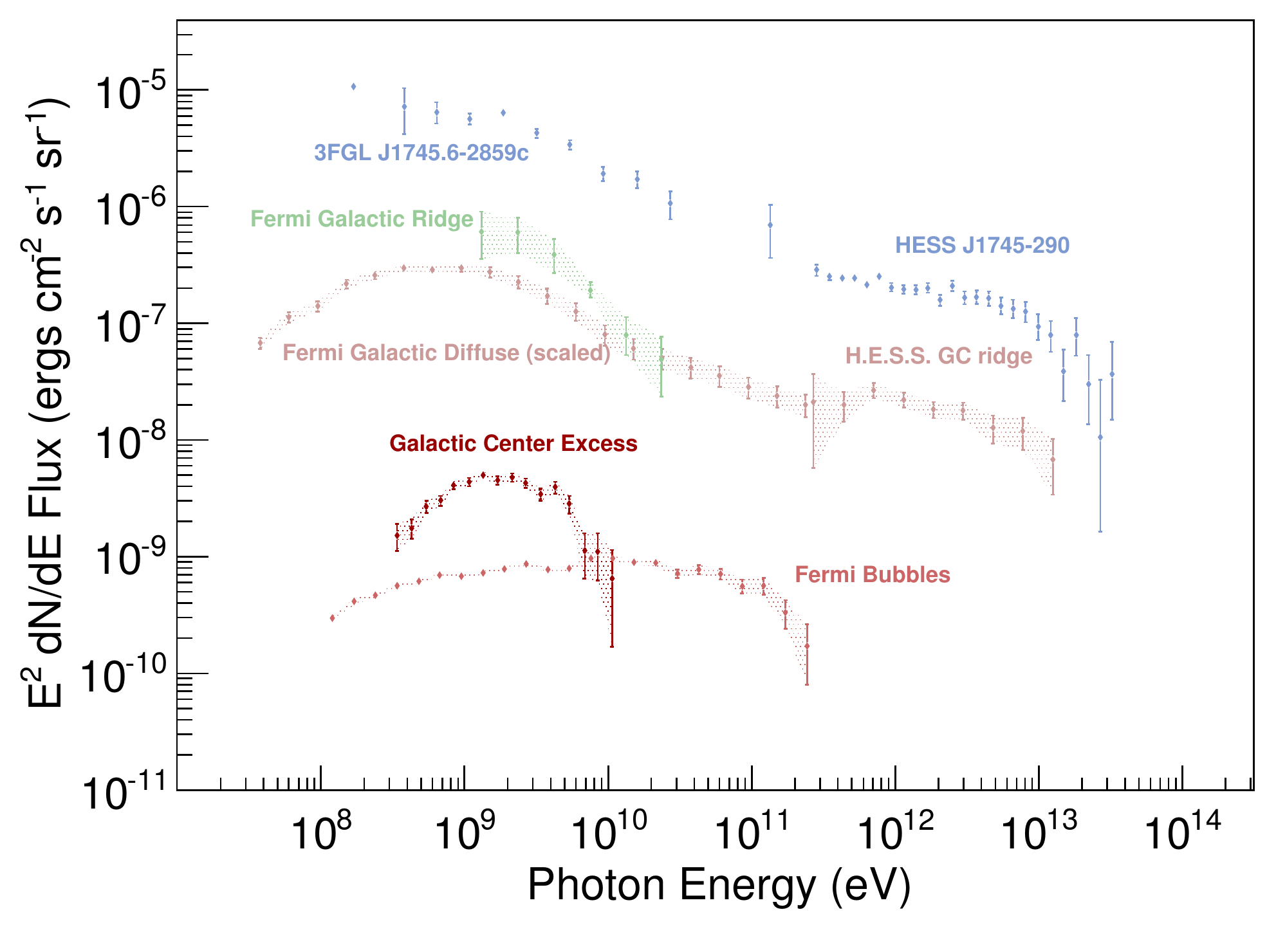}
\caption{Spectral energy distribution for the gamma-ray sources found
  in the inner Galaxy. The most prominent source is the central
  source, coinciding with Sgr A$^\star$ (3FGL\,J1745.6--2859c,
  HESS\,J1745--290) shown in light blue~\citep{gc:hess09,
    Chernyakova}. Next in terms of surface
  brightness is the Galactic center ridge emission (light red) that
  has also been observed at GeV energies (green). Also shown for
  comparison is the {\emph{Fermi}}-LAT Galactic diffuse emission measured in
  the region $|l|<80^\circ$, $|b|<8^\circ$ (labeled {\emph{Fermi
      Galactic Diffuse (scaled)}}). This emission has been scaled up
  by a factor of 5. The factor of 5 is a combination of a factor of 15
  difference in average column densities of neutral interstellar gas
  integrated over the line of sight (assuming
  $X_{CO} = 0.5\times 10^{20}$ molecules cm$^{-2}$ (K km/s)$^{-1}$ for
  the inner region and $X_{CO} = 1.5\times 10^{20}$ molecules
  cm$^{-2}$ (K km/s)$^{-1}$ in the larger region and furthermore
  assuming an HI spin temperature of 150 K, and a dust-to-gas ratio of
  $73 \times 10^{20}$ cm$^{-2}$ mag$^{-1}$ to correct for dark gas)
  and the hadronic emission related to
  the gas being about 30\% of the total emission in the larger
  region~\citep{FermiDiffuse}, resulting in an overall factor of
  5. Also shown is the {\emph{Fermi}}-LAT detected Galactic Center Excess
  (dark red) taken from~\citep{Daylan2014} and the emission from the
  Fermi-Bubbles~\citep{FermiBubbles_Fermi} fitted in a much larger
  region. It should be
  noted, that in a consistent analysis, all components
  would have to be fitted simultaneously (along with all the other
  point and extended sources in the region), while some of the
  publications referred to in the text have done this, not all of the
  publications from which the spectral data points are taken have done
  this for the Galactic center region. } \label{fig::GCSpectra}
\end{figure}

The most prominent source of gamma rays across all energy bands in
this region is coinciding with the kinematic center of the Galaxy and
the strong compact radio source Sgr A$^\star$, the location of the
$4.3 \times 10^6 M_{\odot}$ supermassive black hole. The gamma-ray
source was detected at TeV energies (HESS\,J1745--290) by several
IACTs in the early 2000s~\citep{HESS:gc04, GC:Whipple04, GC:MAGIC06}
and also at GeV energies~\citep{1FGL,2FGL,3FGL,Chernyakova} -- dubbed
3FGL\,J1745.6--2859c. The spectral shape of the GeV-TeV emission does
not favor a dominant dark matter annihilation scenario (see
Figure~\ref{fig::GCSpectra} and ~\citep{Chernyakova,GC:Profumo05}),
although ~\citep{Belikov} and ~\citep{Cembranos} recently suggested a
combination of a powerlaw and a heavy dark matter annihilation
spectrum) and an association with an astrophysical particle
accelerator seems more likely. Detailed studies of the position of the
source with the high-angular resolution of H.E.S.S.\ has provided the
most accurate position to date~\citep{GCPositionHESS}: the source is
within $8'' \pm 9''_{stat} \pm 9''_{sys}$ from the position of Sgr
A$^\star$ with an upper limit on the extension of the source of 1.3'
(95\% confidence level). This measurement rules out the SNR Sgr A East
as a source of the gamma-ray emission, leaving two main objects as
possible astrophysical counterparts: the supermassive black hole Sgr
A$^\star$ or the pulsar wind nebula PWN
G359.95--0.04~\citep{GC:Wang06}. PWNe are known to accelerate
particles (mostly e$^-$ and $e^+$) to beyond TeV energies (see
e.g. the Crab Nebula) and a scenario for the H.E.S.S. spectrum was
presented
by~\citep{GC:Hinton07}. The supermassive black hole itself might also
accelerate particles~\citep{GC:AharonianNeronov05} and the gamma-ray
emission could originate within an O(10) pc zone due to interaction of
runaway protons with the ambient
medium~\citep{GC:AharonianNeronov05apj, GC:Wang09} or by electrons
accelerated in termination shocks driven by strong winds from the
black hole~\citep{AtoyanDermer2004}. With the addition of the
{\emph{Fermi}}-LAT data to lower energies~\citep{Chernyakova} the
spectral shape seems to rule out a single-source leptonic scenario
from G359.95--0.04 and emission related to Sgr A$^\star$ seems the
most likely explanation. A single hadronic scenario for the emission
is still possible, although the diffusion coefficient derived from the
point-like nature of the emission is rather small. Various authors
have employed more complex hybrid or flaring
scenarios~\citep{Chernyakova, Kusunose2012, Guo2014} that can explain
the gamma-ray spectral and morphological properties. In the future CTA
observations might further improve the localisation accuracy of the
gamma-ray source. In addition, a compelling way to relate the emission
to the
black hole has recently been proposed by ~\citep{Abramowski2013}. The
passage of early-type or giant stars approaching the line of sight
will induce pair-production eclipses that will produce a
characteristic time- and energy-dependence in the gamma-ray emission
in the 100--300 GeV band. The projected dimming by a few percent for a
few weeks should be detectable with an instrument such as CTA and
would locate the size of the VHE emitting region to the inner 1000
Schwarzschild radii.

\subsubsection{The Galactic Center ridge} 
Moving further out, H.E.S.S. has reported the detection of diffuse
gamma-ray emission from the Galactic center ridge that is well
correlated with molecular clouds in the central 200 pc of the Milky
way and is confined to the region $|l| < 0.8^\circ$,
$|b| < 0.3^\circ$~\citep{hess:gc_diffuse}. The close correlation with
the interstellar material and the spectral shape (photon index of
$\Gamma = 2.3$) of the emission suggested a hadronic origin in which
what is observed is the result of CRs diffusing from the region. The
spectrum is harder than the spectrum of the Galactic disk and the flux
above 1 TeV is about a factor of 3-9 times higher than that in the
Galactic disk. This led the authors to argue for an additional
population of CRs in the inner 200 pc of the Milky way,
possibly produced by a combination of several SNRs~\citep{hess:gc_diffuse}.
Recently, \citep{YusefZadeh2012} showed that the GeV ({\emph{Fermi}}-LAT)
gamma-ray emission in this region is well correlated with the extended
morphology observed at radio, X-ray and TeV energies and can be
described with a broken power law spectral model with
$\Gamma_1 = 1.8$, $\Gamma_2 = 3.0$, and $E_{\mathrm{break}} = 2.5$GeV. These
authors suggested a non-thermal bremsstrahlung origin of the
broad-band emission from two populations of electrons (one cooled
population responsible for the radio and GeV emission, the other
uncooled population responsible for the TeV emission), although the
magnetic field required for this interpretation is surprisingly low --
O(10 $\mu$G). Also \citep{MaciasGordon2014} concluded that a GeV
component coincident with the H.E.S.S.-detected TeV Galactic ridge
emission is present and that its morphology matches the 20-cm radio
emission. Other hybrid models that employ a combination of hadronic
and leptonic processes to describe the gamma-ray emission in the ridge
are e.g. given in~\citep{Yoast-Hull2014}

\subsubsection{The {\emph{Fermi}}-LAT Galactic Center Excess}
The most tantalising hint for new physics has been detected in the
form of an extended radially symmetric excess in the very center of
our Galaxy. After the original claim by~\citep{HooperGoodenough}, many
studies have confirmed the basic properties~\citep{Daylan2014,
  HooperLinden2011, AbazajianKaplinghat, MaciasGordon2014,
  AbazajianCanac, HooperSlatyer}, although the stability
of these properties with respect to changes in the assumptions about
the Galactic diffuse foreground model is still an open issue. It
should be noted that the signal comprises about 10\% of the total
emission in the inner Galaxy when integrated in the appropriate
region. Nevertheless, many of the alleged properties of the signal
seem to be consistent with a dark matter annihilation origin for a
particle of mass 10--40 GeV for a radially symmetric dark matter
profile following $\rho_{DM} \propto r^{-y}$ with $r$ being the radial
distance to the Galactic center and $\gamma$ in the range $1.1$ to
$1.3$~\citep{HooperReview}. The velocity-averaged annihilation
cross-section needed to explain the emission with a dark matter origin
comes out to a reasonable $\sigma_v = 1.6 \times 10^{-26}$
cm$^3$/s. The signal seems to be radially symmetric and is extended
out to rather large radii ($r \sim 10^\circ$, corresponding to
$\sim 1.5$ kpc). It should, however, be noted, that also astrophysical
sources might be responsible for the emission, most notably CRs
injected in the Galactic center~\citep{CarlsonProfumo2014,
  PetrovicSerpicoZaharijas2014} or a population of unresolved point
sources such as milisecond pulsars
(MSPs)~\citep{AbazajianCanac,Wang2005, Abazajian2011,
  HooperCholis2013, Mirabal2013, CaloreDiMauro2014,
  Cholis2014, Petrovic2014}. MSPs show a similar
spectrum, peaking at $\sim 1-3$GeV and their spatial distribution
might match the excess signal (as suggested by the spatial
distribution of low-mass X-ray
binaries~\citep{VossGilfanov2007,AbazajianKaplinghat}. Determining the
properties of the unresolved population of MSPs relies heavily on
{\emph{Fermi}}-LAT data itself (derived from the $\sim 60$ detected MS pulsars)
which might already constitute a biased sample and might not be
representative for the MSP population in the very center of the
Galaxy. As noted by~\citep{CrockerReview}, DM annihilation would
release energy in annihilation products at a rate of
$\sim 3 \times 10^{37}$ erg/s within the innermost 150 pc of the
Galaxy compared to $10^{39}$ erg/s injected in accelerated hadrons and
$10^{38}$ erg/s in accelerated leptons from star-formation (SNRs,
stellar winds). Energetically, a scenario in which the emission is
related to CR protons or electrons is therefore completely
reasonable~\citep{CarlsonProfumo2014, PetrovicSerpicoZaharijas2014}.

\subsubsection{The {\emph{Fermi}}-LAT Galactic Center Line emission} 
A lot of excitement also revolved around the alleged detection of a
gamma-ray line feature at 130 GeV in the Galactic center data of the
{\emph{Fermi}}-LAT~\citep{Weniger}. While the original publication claimed a
post-trial significance of 3.2$\sigma$ that was later confirmed by
others~\citep{Temple2012, SuFinkbeiner2012a}, the time-evolution of the
signal did not increase as expected for a real signal. Furthermore, an
analysis with updated instrument response functions (\emph{Pass 7
  reprocessed}) further dropped the significance of this signal, so
that at this point the data points towards a statistical fluctuation
rather than a real signal~\citep{Fermi:LinePaper}.

\subsubsection{The {\emph{Fermi}} Bubbles}
Possibly the most surprising discovery in all of high-energy
astrophysics in recent years has been the discovery of the
{\emph{Fermi bubbles}}, large scale structures that extend up to 11 kpc
above and below the Milky way galaxy. The spectrum of these structures
is hard $\sim E^{-1.9 \pm 0.2}$ with an exponential cut off at high
energies at $110 \pm 50$ GeV and also rolling over below 1 GeV. The
emission is rather uniform in intensity across the bubbles (with an
enhancement in the south-east) and the spectrum does not change
throughout the bubbles. The edges are rather sharp
($\sim 3.5^{\circ}$). The bubble spectral shape matches both a
leptonic and a hadronic scenario, although if the WMAP haze is taken
into account the leptonic scenario fits all available data with less
assumptions~\citep[see e.g.][]{MertschSarkar}. Given their position on
the Galactic center, the bubbles seem to originate from that region
(also evidenced by the tighter waist in the Galactic plane). An
obvious candidate is activity in the recent past ($<1$ Myr) of the
SMBH - which could easily supply the energy required to blow the
bubbles ($\sim 10^{56}$ erg). Other scenarios include nuclear
star-formation in the Galactic center region that could in principle
inflates the bubbles if the hot gas, CRs and magnetic fields can be
integrated over sufficiently long
timescales~\citep{CrockerAharonian2011, Crocker2014}. Other
explanations include a jet from the black hole~\citep{Muo2014,
  GuoMathews2012,GuoMathews2012b,Yang2014}, a
spherical outflow from the black hole~\citep{Zubovas2011} or a
sequence of shocks from several accretion events onto the black
hole~\citep{Cheng2011}.

What should be clear from the description in this section is that the
Galactic center region is highly complex and complicated with a large
number of not-yet-understood astrophysical acceleration and diffusion
phenomena. The situation is further complicated by the presence of a
highly structured and extremely bright diffuse gamma-ray background
arising from the interaction of the pool of CRs with dense material in
the inner Galaxy. The whole region is our closest test-bed to study
these phenomena. On top of it, there is the exciting possibility of
new physics (dark matter annihilation) in this region -- however,
before a signal can be claimed, all other astrophysical phenomena have
to be understood to a level where they can be ruled out to contribute
to the alleged dark matter signal.

\subsection{High-energy charged particles in space}

\subsubsection{Particle acceleration}

The two main mechanisms that can accelerate particles in astrophysical
environments are: static electric fields or head-on collisions with
magnetic turbulences. The acceleration in electric fields is readily
understood, however, in astrophysical systems special conditions are
necessary to achieve a large-scale electric field due to the high
conductivity of astrophysical plasmas (the electric fields are
short-circuited by the motion of the free charges). Examples for
astrophysical systems in which large-scale electric fields play a role
are {\emph{unipolar inductors}} (rotating black holes or neutron
stars) or systems with {\emph{magnetic reconnection}} (where locally
regions with opposite orientations of the magnetic field merge).

The second probably more common acceleration mechanism happens in
astrophysical shocks -- where two fluids are compressed by the
compression ratio $r$. In shocks {\emph{Fermi processes}} (head-on
collisions of particles with magnetic turbulences) can accelerate
particles~\citep[for a recent review of shock acceleration see
e.g.][]{Reynolds}. 

Fermi's original idea was to propose a mechanism in a setting in which
(random) collisions with magnetic mirrors between charged particles
and interstellar clouds could result in the acceleration of
particles~\citep{FermiAcceleration}. In the original proposal in which
clouds with random propagation directions were the agents of
collisions, the process was relatively slow and
inefficient. Energy-increasing approaching and energy-decreasing
receding collisions were cancelling each other out and therefore the
energy gains per collision were only second order in the ratio of
cloud velocity $u$ to particle speed $v$, or
$\Delta E/E \sim (u/v)^2$. In the late 1970s several
researchers~\citep{AxfordLeerSkadron1977,Krymsky1977,
  Bell1978,BlandfordOstriker1978} realised independently that in
strong shocks a reference frame existed in which the fluids before and
behind the shocks always converged -- and therefore only approaching
(head-on) collisions which increased the energy of the particle were
happening. These processes were shown to allow for energy gains of
first order in $u/v$, or $\Delta E/E \sim u/v$, with every shock
crossing and therefore a much more rapid acceleration. An important
prediction of the theory is that the accelerated particles exhibit a
number spectrum in momentum space (in the test particle limit in which
the accelerated particles are energetically unimportant) following
$N(p) \sim p^{-3r/(r-1)}$. For strong shocks $r =4$ and therefore
$N(p) \sim p^{-4}$ or $N(E) \sim E^{-2}$.

The second important property of the acceleration process is the
maximum energy the particles can reach. For this property the balance
between energy gain and energy losses during the acceleration become
relevant. Very generally energy can be lost if the particle escapes
from the shock region (if the gyro-radius of the particle exceeds the
shock size), or if the energy losses of the particles
become larger than the energy gains per shock crossing. These
processes lead to an energy spectrum of the accelerated particles that
resembles a powerlaw in energy with an exponential cutoff above some
energy $E_{\mathrm{max}}$~\citep{WebbDruryBiermann1984,Drury1991}. In
general an expression for $E_{\mathrm{max}}$ can be found by equating
the acceleration time in the shock to the lesser of the radiative loss
time or the escape time. As shown by~\citep{Reynolds} the typical
maximum energy achievable is:

\begin{eqnarray}
\label{eq::energymax}
E_{\mathrm{max, age}} \sim 0.5u^2_8 t_3 B_{{\mu}G} \qquad \mathrm{TeV}\\
E_{\mathrm{max, loss}} \sim 100 u_8 B_{{\mu}G}^{-1/2} \qquad \mathrm{TeV}
\end{eqnarray}

With $u_8 = u/10^8$ cm s$^{−1}$, $t_3 = t/1000$ years and $B$ the upstream
magnetic field in $\mu G$.

 Finally (especially for older systems), if escape of
 particles from the acceleration region is the limiting factor, the
 maximum energy should behave as.
 
 \begin{equation}
 E_{\mathrm{max}}(\mathrm{escape}) \sim 10 B_{{\mu}G} \lambda_{17}
   \qquad \mathrm{TeV}
 \end{equation}
 with $\lambda_{17}= \lambda{\mathrm{max}}/10^{17}$ cm. 
 
For typical timescales and shock speeds ($t_3$, and $u_8$ of order 1),
these are energies accessible by ground-based arrays of gamma-ray
telescopes. 

It can be shown, that for any energetically possible source of CRs, a
substantial amount of energy must be contained in the accelerated
particles. Those particles will have a dynamical (non-linear) effect
on the shock which means that the test-particle limit will no longer
be applicable~\citep{MalkovDrury2001}. Most importantly, the pressure of the
accelerated particles will slow down the incoming fluid (in the shock
frame) by diffusing ahead. These non-linear effects will therefore
decrease the compression ratio close to the shock (and thereby make
the spectrum softer i.e. $\Gamma$ larger) but increase the compression
ratio far ahead of the shock (and thereby make the spectrum harder
i.e. $\Gamma$ smaller). The resulting overall particle spectrum would
deviate from the traditional $\Gamma=2$ case both at low energies
(where the spectrum would be softer) and at high energies (where the
spectrum would be harder). A clear test of the theory and the
non-linear effects described here would be a concave spectrum of the
accelerated particles (note that this effect is most relevant for
protons).

\subsubsection{Particle Propagation}
One of the fundamental assumption about the shock acceleration
processes is that particles move diffusively through space with
$ \langle r^2\rangle = 2Dt$ (with $D$ the Diffusion coefficient). $D$
governs the speed of the diffusion process and depends on both the
average magnetic field $B$ and its turbulence $\delta B$ (on
length-scales comparably to the gyroradius $R_g$). In the Bohm limit
-- the slowest possible diffusion -- particles diffuse perpendicular
to magnetic field lines. The mean free path of the particles is thus
given by the gyroradius and the resulting diffusion coefficient can be
estimated as $D \sim \eta R_g c/3$ with
$\eta \approx (\delta B/B)^2$~\citep{StrongMoskalenkoPtuskin2007}. The
gyroradius for particles of unit charge can be written as
$R_{g, pc} \approx 0.001E_{\mathrm{TeV}} / B_{\perp, \mu G}$ and
therefore following~\citep{HintonHofmann}:
\begin{equation}
\langle r^2_{pc} \rangle^{1/2} \approx 0.01(\eta E_{\mathrm{TeV}}
t_{yr}/B_{\perp, \mu G})
\end{equation}
When describing the diffusion of particles in our Galaxy as a whole
one typically comes to the conclusion that the diffusion coefficient
is of the form $D = 10^{28} (E/10\mathrm{GeV})^{\delta}$cm$^2$s$^{-1}$
with $\delta = 0.3-0.7$~\citep{Berezinskii1990}. This energy dependence
can be assumed to be spatially varying and should reflect the local
distribution of magnetic turbulence.
 
\subsubsection{Energy losses}

Energetic particles lose energy via various processes which invariably
leads to the generation of photons all across the electromagnetic
spectrum from radio to high-energy gamma rays. In the gamma-ray energy
band, there are various competing processes that can be used to
explain the observed gamma-ray emission. The main radiation and energy
loss mechanisms all involve the interaction of high-energy charged
particles with magnetic fields (synchrotron emission of both electrons
and protons), photon fields (inverse Compton scattering of electrons
or photo-meson production from protons) or matter (bremsstrahlung, p-p
collisions).

\paragraph{Electron cooling times}

\begin{figure}[htb]
\includegraphics[width=\textwidth]{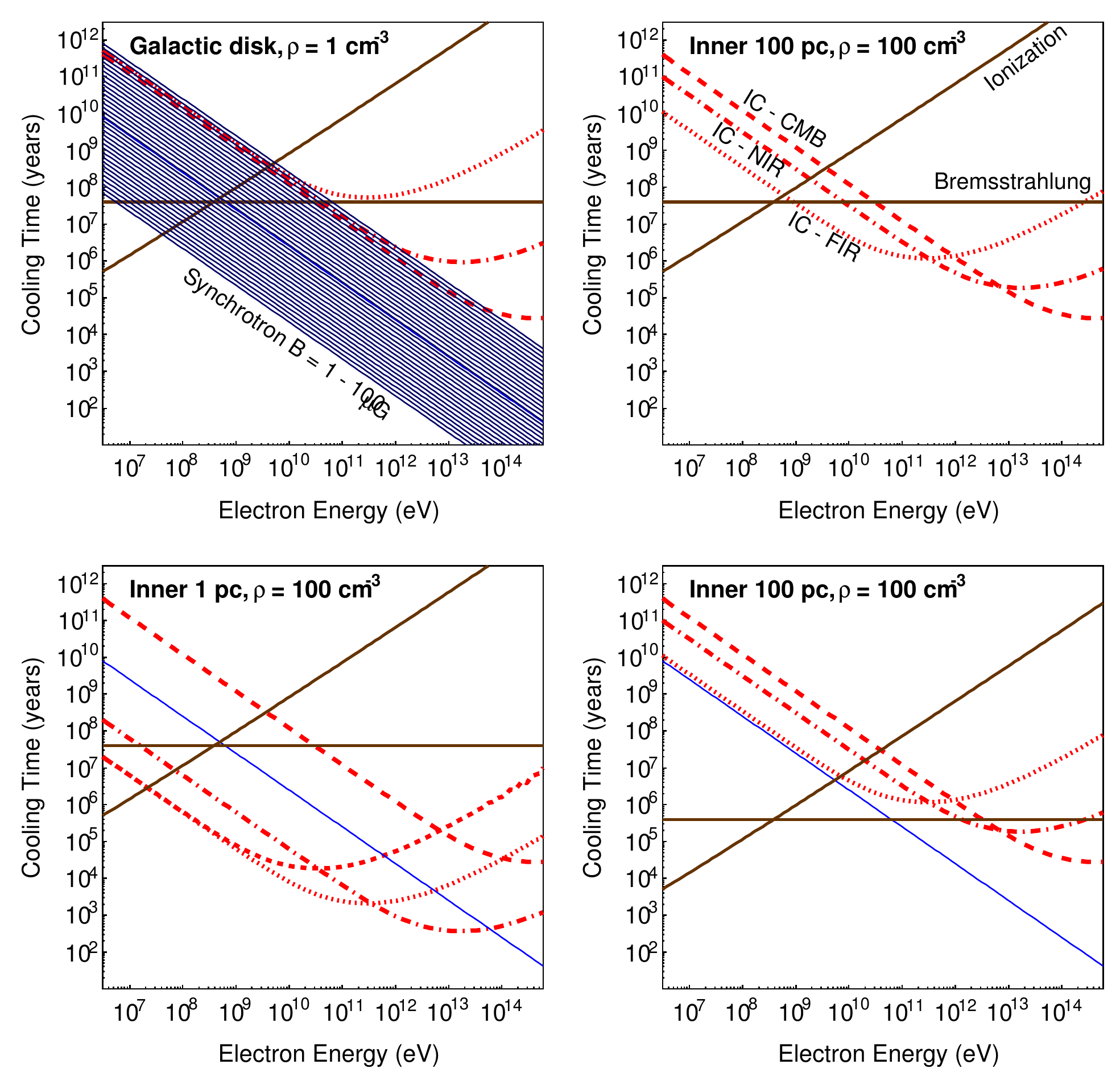}
\caption{Cooling time scales for relativistic electrons for various
  processes in typical Galactic environment. For all plots the energy
  density of the CMB is
  $U_{\mathrm{CMB}} (kT = 2.35 \times 10^{-4} \mathrm{eV}) = 0.26
  \mathrm{eV cm}^{-3}$,
  the matter density is $n_0=1 cm^{-3}$ unless for the bottom right
  plot where $n_0=100 cm^{-3}$ : (Top left): radiation fields in a
  typical Galactic environment
  $U_{\mathrm{NIR}} (0.3 \mathrm{eV}) = 0.2 \mathrm{eV} cm^{-3}$,
  $U_{\mathrm{FIR}} (6\times 10^{-3} \mathrm{eV}) = 0.2 \mathrm{eV
    cm}^{-3}$.
  (Top right): radiation fields in the Galactic disk in the inner 100
  pc from the Galactic center
  ($U_{\mathrm{NIR}} (0.3 \mathrm{eV}) = 9 \mathrm{eV cm}^{-3}$,
  $U_{\mathrm{FIR}} (6\times 10^{-3} \mathrm{eV}) = 1 \mathrm{eV
    cm}^{-3}$).
  (Bottom left): radiation fields in the inner $1 pc^{3}$ from the
  Galactic center
  ($U_{\mathrm{Opt}} (3 \mathrm{eV}) = 5000 \mathrm{eV cm}^{-3}$,
  $U_{\mathrm{NIR}} (0.3 \mathrm{eV}) = 5000 \mathrm{eV cm}^{-3}$,
  $U_{\mathrm{FIR}} (6\times 10^{-3} \mathrm{eV}) = 5000 \mathrm{eV
    cm}^{-3}$).
  Bottom right: like top right (inner 100 pc) with a matter density of
  $n_0=100 \mathrm{cm}^{-3}$.}\label{fig::lossRates}
\end{figure}

Unless in very dense radiation fields electrons in typical Galactic
environments above energies of $\sim 1$GeV lose energy mainly through
synchrotron radiation or (at low magnetic fields) through inverse
Compton scattering. Given the ubiquity of the cosmic microwave
background (CMB), as well as other photon fields in the Universe,
inverse Compton scattering is present with high efficiency over the
whole gamma-ray band. In the Thomson regime inverse Compton cooling
decreases linearly with energy and therefore this process is
particularly important at very high electron energies. 

The expression
for the Inverse Compton scattering cooling time
is~\citep{BlumenthalGould}:

\begin{equation}
\tau_{\mathrm{IC}} = 4 \times 10^8 f _{KN}^{-1}\left( \frac{U_R}{1\mathrm{eV cm}^{-3}} \right)^{-1}
\left( \frac{E_e}{\mathrm{GeV}} \right)^{-1} yr 
\end{equation}

Where $U_R$ is the radiation density of the photon fields and $E_e$ is
the electron energy. In the Thomson regime the loss rate
$dE/dt (=E_e/\tau)$ is therefore proportional to $E^2$ and therefore
Inverse Compton scattering depletes the high-energy part of the
electron spectrum stronger than the low-energy part (resulting in a
steeper electron spectrum $dN/dE$).  For highly energetic electrons
the classical electron cross section gets reduced through relativistic
effects ({\emph{Klein-Nishina}} or KN formula) essentially due to
energy conservation between the up scattered photon and the incident
electron.  The corresponding increase in loss time can be
parameterised~\citep{Moderski2005} by the factor
$f_{KN} \approx (1 + b)^{-1.5} = (1 + 40 E_{e, \mathrm{TeV}} k T_{\mathrm{eV}})^{-1.5}$.
This factor is approximately equal to 1 for small energies (Thomson
regime) and proportional to $ln(E_e)/E_e^2$ for very large energies
with a transition regime in between. Thus in the deep KN regime
$\tau_{\mathrm{IC}} \propto E_e/ln(E_e)$ and therefore the electron spectrum
becomes harder with IC losses.
It is interesting to note, that in the Thomson regime the cooling
time does not depend on the spectrum of target photons but only on the
total radiation density. This is not true for the Klein-Nishina regime.
The expression for synchrotron cooling is very similar to the above
expression for Inverse Compton scattering when one replaces $U_r$ with
$U_B = B^2/(8\pi)$:

\begin{equation}
\tau_{\mathrm{Sync}} = 2.5 \times 10^9 \left( \frac{B}{1\mu G} \right)^{-2}
\left( \frac{E_e}{\mathrm{GeV}} \right)^{-1} yr 
\end{equation}

The cooling time for electrons due to Bremsstrahlung is
energy-independent and depends only on the density of hydrogen atoms
$n_0$:

\begin{equation}
\tau_{\mathrm{Br}} = 4 \times 10^7 \left( \frac{n_0}{1 cm^{-3}} \right)^{-1} \mathrm{yr} 
\end{equation}

\begin{figure}[htb]
\includegraphics[width=\textwidth]{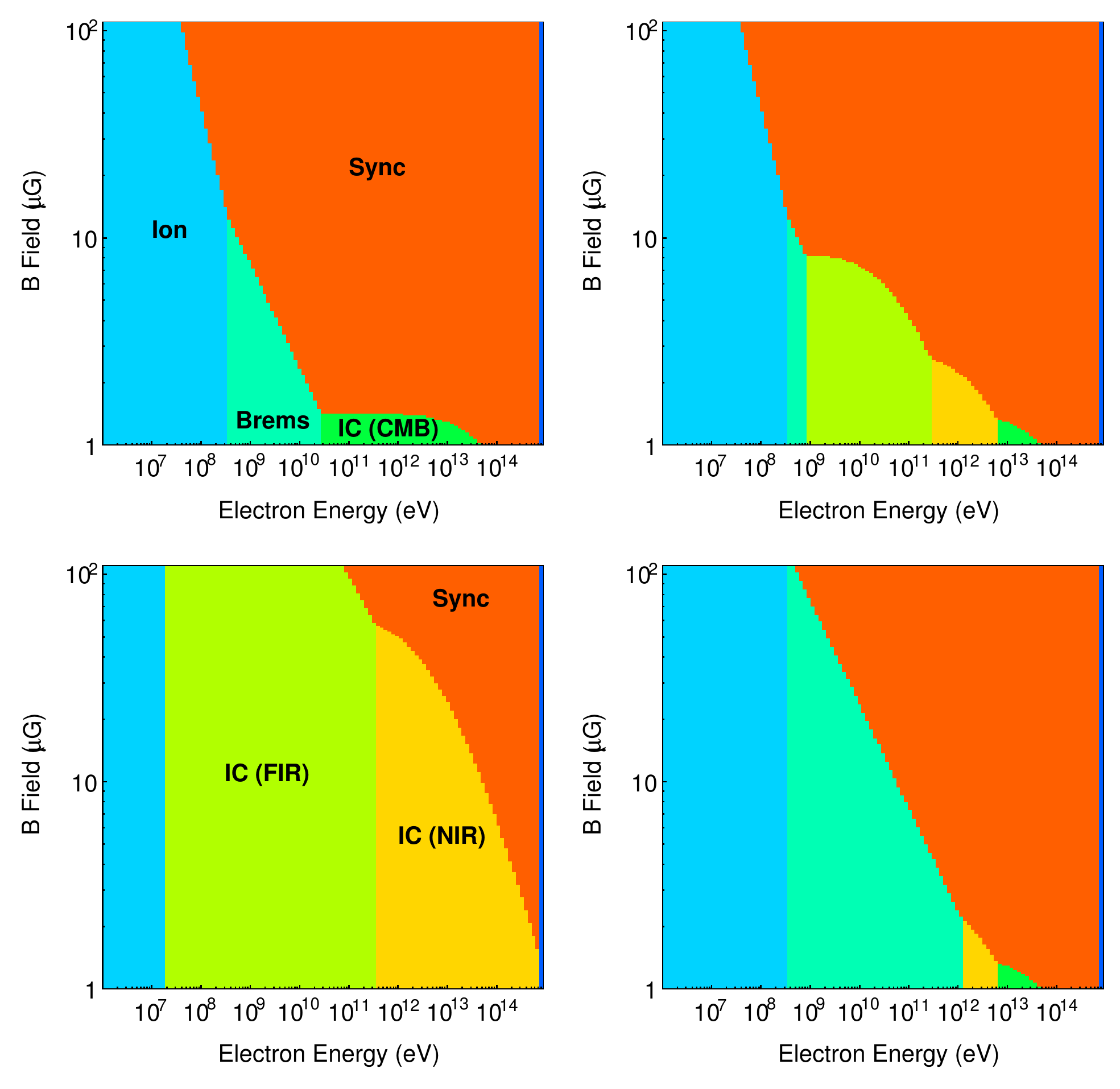}
\caption{Dominant loss processes (i.e. the process with the minimum
  cooling time) for the radiation fields and
  densities used in Figure~\ref{fig::lossRates} as a function of
  magnetic field and as a function of energy. In the blue region
  ionization dominates, in the red region synchrotron emission, light
  green corresponds to Bremsstrahlung, and green and orange to the
  various Inverse Compton components.}\label{fig::dominantCooling}
\end{figure}

The density and location of the interstellar material (mostly consisting
of molecular clouds, HI, and H$_2$ regions) can be measured using
radio, microwave and optical observations~\citep{Cox2005}.  Because the
cooling time is energy-independent Bremsstrahlung does not change the
initial electron energy spectrum - even further, the bremsstrahlung
gamma-ray spectrum simply repeats the parent electron spectrum. In
hydrogen gas below $\sim 350$ MeV ionisation losses dominate over
Bremsstrahlung. The corresponding cooling time
is~\citep{SchleicherBeck2013}:

\begin{equation}
\tau_{\mathrm{Io}} = 4.1 \times 10^9 \left( \frac{E_e}{\mathrm{GeV}} \right) \left(
  (3*ln(E_e/\mathrm{GeV}) + 42.5) \frac{n_0}{1 \mathrm{cm}^{-3}} \right)^{-1}  \mathrm{yr} 
\end{equation}

The electron loss processes are illustrated in
Figure~\ref{fig::dominantCooling} for different Galactic environments
(top left: Galactic disk, top right: inner 100 pc in the Galactic disk
with respect to the Galactic Center, bottom left: inner 1 pc of the
Galaxy, bottom right: like top right but with a significantly higher
matter density). As can be seen, at high energies and high magnetic
fields synchrotron emission is the dominant loss process of
relativistic electrons.

\paragraph{Proton cooling times}

The main cooling mechanism for relativistic protons below $10^{15}$ eV
in a hydrogen (target) medium is the inelastic scattering resulting in
the production of secondary pions ($\pi$), eta ($\eta$), kaons and
hyperons. While nearly 100\% of neutral pions ($\pi^0$s) decay into two
photons, the branching ratio of $\eta \rightarrow \gamma\gamma$ is only
40\%.  Similar to the case of Bremsstrahlung the
cooling time above proton energies of $E_p = 1$GeV is almost
independent of energy (a result of the fact that the total pp
scattering cross section $\sigma_{pp}$ is almost
energy-independent),
 The cooling time can be given as:

\begin{equation} 
\tau_{\mathrm{pp}} = (n_0)^{-1} (\sigma_{\mathrm{pp}} f
  c)^{-1} = 5.3 \times 10^7 \left( \frac{n_0}{1 \mathrm{cm}^{-3}}
  \right)^{-1} \mathrm{yr} 
\end{equation}

therefore the initial (acceleration) energy spectrum of the protons
remains unaffected. At higher proton energies and extragalactic
environments with low hydrogen densities proton synchrotron and
photo-pion production becomes important.


\bibliographystyle{ar-style5.bst}
\bibliography{funk_AR_gammaAstrophysics}

%
%
\end{document}